\documentclass[conference]{IEEEtran}
\usepackage{filecontents,lipsum}
\usepackage[noadjust]{cite}
\usepackage{multicol}
\usepackage{setspace}
\usepackage{indentfirst}
\usepackage{makecell}
\usepackage{tocloft}
\usepackage{doi}
\usepackage{hyperref}
\usepackage{breakurl}
\usepackage{url}
\usepackage{tocbibind}
\usepackage{graphicx}
\usepackage{float}
\usepackage[skip=5pt]{caption}
\usepackage{array}
\usepackage{algorithmic}
\usepackage{amsmath}
\usepackage{amssymb}
\usepackage{tabularx}
\usepackage{longtable}
\usepackage{xcolor}
\hypersetup{hidelinks}
\usepackage{setspace}
\usepackage{float}
\usepackage{amsmath,amssymb,amsfonts}
\usepackage{algorithmic}
\usepackage{textcomp}
\IEEEoverridecommandlockouts
\def\BibTeX{{\rm B\kern-.05em{\sc i\kern-.025em b}\kern-.08em
    T\kern-.1667em\lower.7ex\hbox{E}\kern-.125emX}}

\begin{document}

    \title{AN EMPIRICAL EVALUATION OF THE IMPLEMENTATION OF THE CALIFORNIA CONSUMER PRIVACY ACT (CCPA)\\}

\author{\IEEEauthorblockN{Trong Nguyen}
\IEEEauthorblockA{\textit{California State Polytechnic University, Pomona} \\
3801 West Temple Avenue
Pomona, California 91768 \\
trongnguyen@cpp.edu}
\and
\IEEEauthorblockN{ Mohammad Husain}
\IEEEauthorblockA{\textit{California State Polytechnic University, Pomona} \\
3801 West Temple Avenue
Pomona, California 91768 \\
mihusain@cpp.edu}
}

\maketitle

\pagenumbering{roman}
% \input{title}\newpage
% \begin{singlespace}
% \input{signature}\newpage
% \end{singlespace}

\begin{center}

\section*{\centering ACKNOWLEDGMENTS}
\addcontentsline{toc}{section}{Acknowledgments}
\end{center}
I want to start by thanking families. Without their support, none of my achievements could have been
possible. Next, I would like to thank you National Science Foundation, for providing support through
CyberCorps Scholarship for Service (SFS) program and Dr. Momhammad Husain’s mentorship, teaching, and guidance. I also want
to acknowledge my committee member Dr. TingTing Chen, for teaching me valuable cyber security skills.
Lastly, I appreciate my friends and professors who have supported me throughout my education.\newpage

\section*{\centering ABSTRACT}
\addcontentsline{toc}{section}{Abstract}
On January 1, 2020, California passed the California Consumer Privacy Act ( CCPA ) by more than 56\% voters that intended to enhance privacy rights and consumer protection for residents of California, United States. Since then, more conditions have been added to the Act to support consumers' privacy \cite{ccpanewrights}. In addition, two years after the first effective day of CCPA, consumers have seen California organizations apply approaches to adapt to CCPA. 

Many organizations quickly upgrade their policy to comply with the legislation and create effective platforms such as data portals that allow consumers to exercise their privacy rights. However, on the other hand, we still noticed aspects of CCPA being absent on some websites. Also, we learned that there exists a technique that a social media company may use to minimize the effect of CCPA on their data collecting businesses. Additionally, we found no prior evaluation of the CCPA implementation in organizations. Therefore, the convergence of the regulatory landscape and the organization's privacy policy needs to be studied. For example, it is feasible to measure consumers' effortlessness in exercising their privacy rights. In addition, when exercising data rights, how challenging consumers may have must be assessed. 

Consequently, this paper was about an empirical evaluation of the implementation of the California Consumer Privacy Act. The report includes the evaluations of the following industries: Social media, financial institutions, mortgages, healthcare providers, and academic institutions. Because of resource constraints, we would choose only the most prominent companies from each category. Our approach was to set up a criteria table constructed from the CCPA Act and then use that table as a checklist while reviewing a company's privacy notice. To evaluate effortlessness in exercising CCPA rights, we set up dummy accounts to generate fiction data, perform data requests such as accessing or deleting, and observe organizations' responses. We used our real accounts for the services that we could not create fake accounts, such as bank accounts but applied obfuscation when including data in the report.

Finally, we concluded this paper with an online tool application design that verifies the CCPA implementation. Upon completion, the application would be free to use so consumers can quickly inspect a website for CCPA compliance. Additionally, it is an advising tool that a website admin can utilize to enhance CCPA compliance for their website. The conjunction of this empirical report and a practical application function as a stimulus to promote CCPA implementation in organizations and deliver awareness to consumers about privacy rights they can demand. 
\newpage

% \renewcommand{\cftsecleader}{\cftdotfill{\cftdotsep}}

% \renewcommand{\contentsname}{Table of Contents}

% \renewcommand{\cfttoctitlefont}{\hfil \Large \bfseries}

% \addtocontents{toc}{\protect\setcounter{tocdepth}{-1}}
% \tableofcontents\newpage
% \addtocontents{toc}{\protect\setcounter{tocdepth}{3}}

% \renewcommand{\cftlottitlefont}{\hfil \Large \bfseries}
% \listoftables\newpage

% \renewcommand{\cftloftitlefont}{\hfil \Large \bfseries}
% \listoffigures\newpage

\pagenumbering{arabic}
\section{Introduction}
\subsection{Moltivation}
Recently, computer and cyber-security communities have noticed many data leak incidents that invade users' privacy \cite{topsecuritybreaches}. Courts have summoned social media companies such as Facebook for testification due to these incidents \cite{facebookdataleak}. Unfortunately, these incidents still keep happening. Social media companies, bank institutions, and healthcare providers have become more influential in the big data era because they can obtain users' data on a larger scale \cite{dangerofbigdata}. Furthermore, every time users visit websites, their data such as geo-location, IP address, email, and phone number will be collected for advertising and profit-making purposes \cite{facebookmakemoney}. Notably, not every data is collected with a consumer's consent \cite{trackwithoutconsent}. In the past, consumers had less control over their data due to the lack of consumer privacy regulation. However, the situation now differs; thanks to the most recent legislation, California has reinforced consumers with a new registration that targets the data collecting activity from companies. 

Understanding the urgent that consumers must have more control over their data, California Authorities have recently passed California Consumer Privacy Act (CCPA) to enhance consumer privacy \cite{ccpawebsite}. The Act is a guideline that mandates California organizations to revise their privacy policies and internal processes to improve user privacy and deliver users better rights over the data. Specifically, the Act requires companies who collect users' data to be transparent on categories and personal information by disclosing them properly. It also informs consumers that they can request companies regarding their data, such as accessing and deleting personal data. Accordingly, many companies have adapted quickly by launching platforms or applying automation processes for consumers to exercise their rights with less effort. Unfortunately, there is still hesitation that some organizations still get in complying with CCPA because the majority of their profit is from sharing consumer data \cite{cccpa-affect-businesses}. For that reason, this paper includes experimental data and reports that demonstrate the current status of CCPA implementation. We emphasize that delaying CCPA implementation can cause monetary damage and reputation loss.

The unawareness of CCPA  can negatively affect companies and consumers. Therefore, besides an empirical report on CCPA implementation, we believe it is vital to have an efficient method for verifying  CCPA on a website instead of manually checking by reading through lengthy pages of privacy notices. Therefore, this paper proposes an online application that will utilize web automation technology \cite{webcralwer} to verify CCPA on websites quickly. In addition, the tool can also advise website administrators to modify their website to comply with CCPA. 
\subsection{California Consumer Privacy Act }
As California residents, we can request that businesses reveal what information they collect
about us. Users have a right to be notified by the time businesses collect their data
Personal information is: Name, social security number, email addresses, records of products
purchased, internet browsing history, geo-location data, fingerprints, and inferences.
\begin{itemize}
    \item Business that CCPA applies to:
        \begin{itemize}
        \item Have a gross annual revenue of over 25 million dollars.
        \item Buy, receive, or sell the personal information of 50,000 or more California residents,
households or devices - Derive 50 percent or more of their annual revenue from selling California
residents' personal information.
        \end{itemize}
    \item Right of consumers:
\begin{itemize}
    \item Have notice of collection: A notice of collection needs to be provided
at or before the business collects personal information.
    \item Right to opt-out of the sale
    \begin{itemize}
        \item There should be a "Do Not Sell My Personal Information" link on the website.
        \item Link has to be working - Right to know personal information that is being collected.
        \item There should be at least two methods, and one of them has to be a toll-free phone number.
        \item There should be an online form on the website to submit the personal information request or
an email address to submit the request if the business operates exclusively online. 
    \end{itemize}
    \item Request to delete personal information.
    \begin{itemize}
         \item Business must have at least two methods for users to submit a request: a toll-free number,
email address, website form, or hard copy form.
        \item Business cannot make consumers create an account just to submit a deletion request.
    \end{itemize} 
    \end{itemize}
\end{itemize}

\subsection{Company Websites Covered In The Research }
This project covers famous companies' websites from the following categories: social media, e-commerce, financial institution, health care, and mortgage companies. First, we looked at popular social media platforms, including Facebook and Google YouTube. In the e-commerce industry, we evaluated Amazon's website. Next, we evaluated  Bank Of America, Wells Fargo, and Chase as the financial institutions. Next, we evaluate Kaiser, Health Net, and Oscar in the healthcare industry. Then the mortgage companies included in our research are Quicken Loans, United Shore Financial, and Freedom Mortgage. Finally, the academic institution category included California State Polytechnic University. We noticed that some companies have multiple websites corresponding to their services, so we only considered the site with the most traffic. For example, Kaiser Permanente has healthy.kaiserpermanente.org as its most popular site for California users to manage their healthcare plans. Instagram is a service under Facebook, and Youtube is a service under Google, so these websites follow the privacy policy provided by their parent companies. 
\subsection{Web Crawler}

The primary technology we used to implement the features is web crawling, Java Selenium. Web crawling is a technology that indexes information on the page using bots known as crawlers. Crawling is similar to what search engines do that will see a web page as a whole and index it. When a bot crawls a website, it goes through every page and link until it reaches the last component of the website. Thus, we used a web crawler to detect text fields related to CCPA criteria; then, we piped the crawled data through some business logic components to determine whether the website complies with CCPA. After that, the front-end user interface employed the modern JavaScript framework VueJS \cite{vuejs} to render and present the result to consumers. 
\subsection{Review of Existing CCPA Related Online Tools }
\label{subsec: Online tool related to CCPA Compliance }
First of all, we performed literature research to see if a similar application exists in the market. The result told us that there were only two CCPA-associated applications, but they were made for other purposes rather than verifying CCPA criteria. The detail is as follows.
\subsubsection{Consumer Privacy Interactive Tool from State of California Department of Justice}
Consumers can use this tool \cite{privacyinteractivetool} to draft a notice sent to businesses that do not post an easy-to-find "Do Not Sell My Personal information." First,  consumers will respond to questions to determine if the targeted company is required to post a "Do Not Sell My Information" notice. Then consumers will fill in information such as the requestor's name and business name. Then the tool will forge a compliance notice and send it to the business that is missing notice of collection. 

\textbf{Limitations}: This application's features are different from our application's features. Generally, this application only serves as a notice drafting tool. It does not verify CCPA criteria on a company's website.

\subsubsection{Cookiebot}
\label{subsec:Cookiebot}
Cookiebot is a free trial tool \cite{cookiebot} to retrieve a report about which data a website is collecting and which third parties they share the data with. The logic behind this tool is to inspect cookies or trackers used by a web browser when visiting a website. Cookies can expose the type of data and the party responsible for that data collecting. A free trial account can scan up to 1 domain and 100 sub-pages. The premium account can browse from \$10 to \$41 per domain from 500 to 5000 sub-pages.

\textbf{Limitations}: This application scans the cookies to determine the collected data and related parties. However, about CCPA compliance checking, this application does not have a feature to verify CCPA criteria. Notably, it cannot check for collection notice and consumer data rights. In addition, the free version is limited to working with websites that have under 100 sub-pages. In contrast, our application will provide all functionalities at no cost that help promote  CCPA. Instead of scanning a website's cookies, our tool will scan the actual text content of the website and compare the data against CCPA criteria.

\newlength{\mycolwidth}
\setlength{\mycolwidth}{\dimexpr \textwidth/4 - 2\tabcolsep}%
\section{Literature survey}
\label{sec:lit_survey}

\subsection{Evaluation of CCPA Implementation}
This research used a criteria table constructed from the Act to create a criteria checklist. Then we used that checklist to verify CCPA compliance on companies' websites. However, the outcome would depend solely on inspecting the website's privacy notice - the CCPA section that companies post on their websites. For that reason, we could not verify some criteria related to internal company processes from our end, such as questions about if the company has trained its employees about CCPA.
\subsubsection{Social Media and E-Commerce}
Based on a ranking report of websites  \cite{topwebsites}, we retrieved a list of the most popular social media websites to perform CCPA checking. These websites are ranked based on the number of page visits. We figured out that all these social websites in this research provided users with platforms for consumers to exercise their rights to access or delete data. However, none of these websites allow users to opt-out data selling because they claim that they do not sell user data. The experiment which requests to access data from these websites gives us our collected data in JSON format\cite{json}. JSON is portable so that users can port their data to other platforms.  \cite{facebookwebsite}\cite{googlewebsite}\cite{amazonwebsite}.

We noticed that these websites were missing data collection notices which should be shown when consumers visit those websites. As a result, these websites do not inform users if their data has been collected unless consumers go to the privacy notice page. As one of the CCPA requirements, organizations should immediately notify consumers if the website contains their data. The following is the outcome of reviewing the CCPA privacy notice on the experimental websites using the CCPA criteria table. \\

\clearpage
\onecolumn

\begin{table}[H]
\caption{CCPA thresholds table} 
\begin{tabularx}{\textwidth}{ 
    | >{\raggedleft\arraybackslash}X 
    | >{\centering\arraybackslash}X
    | >{\centering\arraybackslash}X
    | >{\centering\arraybackslash}X
    | >{\centering\arraybackslash}X
    | >{\centering\arraybackslash}X|}
   
\hline
 &Google.com & Youtube.com & Facebook.com & Amazon.com & Instagram.com\\
\hline\hline
Have a gross annual revenue of over 25 million ?  & Yes & Yes & Yes & Yes & yes\\
\hline
Buy, receive or sell personal information of 50,000 or more California resident's households, devices.
& already met by the first threshold  
& already met by the first threshold 
& already met by the first threshold 
& already met by the first threshold 
& already met by the first threshold \\
\hline
Derive 50 \% or more of their annual revenue from selling California residents's personal information.
& already met by the first threshold  
& already met by the first threshold 
& already met by the first threshold 
& already met by the first threshold 
& already met by the first threshold \\
\hline
\end{tabularx}
\end{table}

\begin{center}
    \begin{longtable}{|p{1.8in}|p{1in}|p{1in}|p{1in}|}
    \caption{CCPA criteria table for social media} \\
\hline 
\multicolumn{1}{|c|}{\textbf{}} & \multicolumn{1}{c|}{\textbf{Google Youtube}} & \multicolumn{1}{c|}{\textbf{Facebook Instagram }} & \multicolumn{1}{c|}{\textbf{Amazon}}\\ \hline 
\endfirsthead

\multicolumn{3}{c}%
{{\mdseries \tablename\ \thetable{}:  Continued from previous page}} \\
\hline \multicolumn{1}{|c|}{\textbf{}} & \multicolumn{1}{c|}{\textbf{Google Youtube}} & \multicolumn{1}{c|}{\textbf{Facebook Instagram }} & \multicolumn{1}{c|}{\textbf{Amazon}}\\ \hline 
\endhead

\hline \multicolumn{3}{|r|}{{Continued on next page}} \\ \hline
\endfoot
\hline \hline
\endlastfoot
    \hline
    The company informs consumers what categories they have collected ?
    & Yes & Yes & Yes \\
    \hline
    Does the company inform consumers which specific pieces of personal information they have collected ? 
    & Yes & Yes & Yes \\
    \hline 
    The company verified consumers' identity who requested to access or delete their personal information ? 
    & Yes & Yes & Yes \\
    \hline
    Does the company inform consumers before the point of collection ? 
    & No notice of collecting on the home page 
    & No notice of collecting on the home page 
    & No notice of collecting on the home page  \\
    \hline
    Does the company deliver information to consumers free of charge within 45 days ?
    & Yes & Yes & Yes \\ 
    \hline
    The information the company has to deliver is portable and in a usable readable format that allows consumers to transmit the information to other entities without hindrance ?
    & Yes & Yes & Yes \\ 
    \hline
    The company delete personal information when a consumer request it.
    & Yes & Yes, but users have to wait 30 days to have all data deleted & Yes \\ 
    \hline
    The company has to create a process and to identify individuals responsible for a consumer to opt-out and, therefore, not selling their data to third parties in response. 
    & No, the company says they do not sell customer data  
    & No, the company says they do not sell customer data  
    & No, the company says they do not sell customer data  \\ 
    \hline
    By default, the company should not sell consumers' personal information when they are between 13 and 16 years old, but there has to be a process to allow them to opt-in.
    & Yes & Yes & Yes \\
    \hline
    The company has to provide consumers the right to equal services and prices. 
    & Yes & Yes & Yes \\
    \hline
    The company has to make available two or more designated methods for the consumer to request their information, including, at a minimum, a toll-free telephone number and website address (if the business maintains a website).
    & Yes & Yes & Yes \\
    \hline
    The company has to train and inform dedicated personnel to process new requests to exercise privacy rights properly.
    & Cannot be verified
    & Cannot be verified 
    & Cannot be verified \\ 
    \hline
    The company has to disclose the consumer's rights to request the deletion of their personal information.
    & Yes & Yes & Yes \\
    \hline
    Suppose the company sells consumers' personal information. In that case, it has to inform your customers that their information may be sold and that they have the "right to opt-out" of selling their personal information.
    & No, the company says they do not sell data 
    & No, the company says they do not sell data
    & No, the company says they do not sell data \\
    \hline
    If the company offers financial incentives for the collection, the sale, or the deletion of personal information, you need to disclose those financial incentives to your consumers.
    & No, the company says they do not sell data 
    & No, the company says they do not sell data
    & No, the company says they do not sell data \\
    \hline
    The homepage website has to include a link to inform consumers that they have the right to opt-out of their information sale.
    & No, the company says they do not sell data 
    & No, the company says they do not sell data
    & No, the company says they do not sell data \\
    \hline
    In its online privacy policy, the company has to disclose a description of consumers' rights and the categories of consumers' personal information collected and/or sold in the preceding 12 months.
    & Yes & Yes & Yes \\
    \hline

    \end{longtable}
    \end{center}

\clearpage
\twocolumn
\subsubsection{Financial Institution}
The financial institution  included in this research Bank Of America (BoA) \cite{boawebsite}, Wells Fargo \cite{wellsfargowebsite}, Chase \cite{chasewebsite} . We only had access to BoA and Chase data portals with our bank accounts. For Wells Fargo, because we do not have an account with them, the review result is solely based on examining the bank privacy notice. We made a data access request to BoA and Chase, but only Chase gave us the data report. BoA did not fulfill our request because our account falls under a consumer account. They claimed that consumer account is exempted from CCPA. Detail of the BoA response is included in the next paper section. 
BoA might use an exemption of CCPA under Gramm-Leach-Bliley Act (GLBA) \cite{glba}. CCPA does not generally apply to personal information collected, processed, sold, or disclosed pursuant to the federal or its implementing regulations. Although data regulated by the GLBA is not generally subject to the CCPA, CCPA still applies to this data with respect to consumers' private of action. For example, the CCPA allows California residents to sue an organization if certain data elements are not encrypted when a data breach occurs. On both Bank Of America and Chase websites,  no collection notice was available when we visited their homepage.

\clearpage
\onecolumn

\begin{center}

\begin{longtable}[c]{|p{\mycolwidth}|*{3}{p{\mycolwidth}|}}
\caption{CCPA criteria table for financial institutions} \\
\hline 
\multicolumn{1}{|c|}{\textbf{}} & \multicolumn{1}{c|}{\textbf{Bank Of America}} & \multicolumn{1}{c|}{\textbf{Wells Fargo}} & \multicolumn{1}{c|}{\textbf{Chase}}\\ \hline 
\endfirsthead

\multicolumn{3}{c}%
{{\mdseries \tablename\ \thetable{}: Continued from previous page}} \\
\hline 
\multicolumn{1}{|c|}{\textbf{}} & \multicolumn{1}{c|}{\textbf{Bank Of America}} & \multicolumn{1}{c|}{\textbf{Wells Fargo}} & \multicolumn{1}{c|}{\textbf{Chase}}\\ \hline 
\endhead

\hline \multicolumn{3}{|r|}{{Continued on next page}} \\ \hline
\endfoot
\hline \hline
\endlastfoot
    \hline
    The company informs consumers what categories they have collected ?
    & Yes & Yes & Yes \\
    \hline
    Does the company inform consumers which specific pieces of personal information they have collected ? 
    & Yes & Yes & Yes \\
    \hline 
    The company verified consumers' identity who requested to access or delete their personal information ? 
    & Yes & Yes & Yes \\
    \hline
    Does the company inform consumers before the point of collection ? 
    & No notice of collection on the homepage
    & No notice of collection on the homepage
    & No notice of collection on the homepage \\
    \hline
    Does the company deliver information to consumers free of charge within 45 days ?
    & Yes & Yes & Yes \\ 
    \hline
    The information the company has to deliver is portable and in a usable readable format that allows consumers to transmit the information to other entities without hindrance ?
    & The company did not present the data explaining that the account that made the request is a consumer account & No account to test & Yes \\ 
    \hline
    The company delete personal information when a consumer request it.
    & Unable to verify  & Unable to verify  & Unable to verify  \\ 
    \hline
    The company has to create a process and to identify individuals responsible for consumers to opt-out and, therefore, not selling their data to third parties in response.
    & No, Company says they do not sell customer data  
    & No, Company says they do not sell customer data  
    & No, Company says they do not sell customer data \\ 
    \hline
    By default, the company should not sell consumers' personal information when they are between 13 and 16 years old, but there has to be a process to allow them to opt-in.
    & Yes  & Yes  & Yes  \\
    \hline
    The company has to provide consumers the right to equal services and prices. 
    & Yes & Yes & Yes \\
    \hline
    The company has to make available two or more designated methods for the consumer to request their information, including, at a minimum, a toll-free telephone number and website address (if the business maintains a website).
    & Yes & Yes & Yes \\
    \hline
    The company must train and inform dedicated personnel to properly process new requests to exercise privacy rights.
    & Unable to verify 
    & Unable to verify
    & Unable to verify\\ 
    \hline
    The company has to disclose the consumer's rights to request the deletion of their personal information.
    & Yes & Yes & Yes \\
    \hline
    Suppose the company sells consumers' personal information. In that case, it has to inform its customers that their information may be sold and that they have the "right to opt-out" of selling their personal information.
    & No, the company says they do not sell data 
    & No, the company says they do not sell data
    & No, the company says they do not sell data \\
    \hline
    If the company offers financial incentives for the collection, the sale, or the deletion of personal information, you need to disclose those financial incentives to your consumers.
    & No, the company says they do not sell data 
    & No, the company says they do not sell data
    & No, the company says they do not sell data \\
    \hline
    The homepage website has to include a link to inform consumers that they have the right to opt-out of their information sale.
    & No, the company says they do not sell data 
    & No, the company says they do not sell data
    & Chase allows users to opt-out of data sharing, but the company also says they do not sell data \\
    \hline
    In its online privacy policy, the company has to disclose a description of consumers' rights and the categories of consumers' personal information collected and/or sold in the preceding 12 months.
    & Yes & Yes & Yes \\
    \hline

    \end{longtable}
    \end{center}

\begin{twocolumn}
\subsubsection{Mortgage Industry}
In this research, we also evaluate mortgage companies for CCPA implementation. Three of the largest mortgage companies, Quicken Loans \cite{quickenloanwebsite}, UWM \cite{unitedshorenwebsite}, and Freedom Mortgage \cite{freedommortgagewebsite}, all mention CCPA on their websites. However, there was no notice of collection when visiting the website. In addition, we do not have accounts to test accessing/deleting data requests.
We generated the results table below based on reviewing the privacy notice on company websites. 
\end{twocolumn}

\clearpage
\onecolumn
\begin{center}
\begin{longtable}[c]{|p{\mycolwidth}|*{3}{p{\mycolwidth}|}}
\caption{CCPA criteria table for mortgage companies} \label{ccpa:long}\\
\hline 
\multicolumn{1}{|c|}{\textbf{}} & \multicolumn{1}{c|}{\textbf{Quicken Loans}} & \multicolumn{1}{c|}{\textbf{UWM}} & \multicolumn{1}{c|}{\textbf{Freedom Mortgage }}\\ \hline 
\endfirsthead

\multicolumn{3}{c}%
{{\mdseries \tablename\ \thetable{}: Continued from previous page}} \\
\hline 
\multicolumn{1}{|c|}{\textbf{}} & \multicolumn{1}{c|}{\textbf{Quicken Loans}} & \multicolumn{1}{c|}{\textbf{UWM}} & \multicolumn{1}{c|}{\textbf{Freedom Mortgage }}\\ \hline 
\endhead

\hline \multicolumn{3}{|r|}{{Continued on next page}} \\ \hline
\endfoot
\hline \hline
\endlastfoot
    \hline
    The company informs consumers what categories they have collected ?
    & Yes & Yes & Yes \\
    \hline
    Does the company inform consumers which specific pieces of personal information they have collected ? 
    & Yes & Yes & Yes \\
    \hline 
    The company verified consumers' identity who requested to access or delete their personal information? 
    & Yes & Yes & Yes \\
    \hline
    Does the company inform consumers before the point of collection ? 
    & No notice of collection when visiting the website 
    & No notice of collection when visiting the website 
    & No notice of collection when visiting the website  \\
    \hline
    Does the company deliver information to consumers free of charge within 45 days ?
    & Do not have an account to test  & Do not have an account to test  & Do not have an account to test  \\ 
    \hline
    The information the company has to deliver is portable and in a usable readable format that allows consumers to transmit the information to other entities without hindrance ?
    & Do not have an account to test  & Do not have an account to test  & Do not have an account to test  \\ 
    \hline
    The company delete personal information when a consumer request it.
    & Do not have an account to test  & Do not have an account to test  & Do not have an account to test  \\ 
    \hline
    The company has to create a process and to identify individuals responsible for consumers to  opt-out and, therefore, not selling their data to third parties in response.
    & Yes, but the company says they do not sell customer's information  
    & Yes, but the company says they do not sell customer's information  
    & No,  but the company says they do not sell customer's information \\ 
    \hline
    By default, the company should not sell consumers' personal information when they are between 13 and 16 years old, but there has to be a process to allow them to opt-in.
    & Not mentioned on the website  & Not mentioned on the website  & Not mentioned on the website   \\
    \hline
    The company has to provide consumers the right to equal services and prices. 
    & Yes & Yes & Yes \\
    \hline
    The company has to make available two or more designated methods for the consumer to request their information, including, at a minimum, a toll-free telephone number and website address (if the business maintains a website).
    & Yes & Yes & Yes \\
    \hline
    The company has to train and inform dedicated personnel to process new requests to exercise privacy rights properly.
    & Unable to verify 
    & Unable to verify
    & Unable to verify\\ 
    \hline
    The company has to disclose the consumer's rights to request the deletion of their personal information.
    & Yes & Yes & Yes \\
    \hline
    Suppose the company sells consumers' personal information. In that case, it has to inform its customers that their information may be sold and that they have the "right to opt-out" of selling their personal information.
    & No, the company says they do not sell data 
    & No, the company says they do not sell data
    & No, the company says they do not sell data \\
    \hline
    If the company offers financial incentives for the collection, the sale, or the deletion of personal information, you need to disclose those financial incentives to your consumers.
    & No, the company says they do not sell data 
    & No, the company says they do not sell data
    & No, the company says they do not sell data \\
    \hline
    The homepage website has to include a link to inform consumers that they have the right to opt-out of their information sale.
    & No, the company says they do not sell data 
    & No, the company says they do not sell data
    & Chase allows users to opt-out of data sharing, but the company also says they do not sell data \\
    \hline
    The company has to disclose in its online privacy policy a description of consumer's rights and the categories of the consumer's personal information collected and/or sold in the preceding 12 months.
    & Yes & Yes & Yes \\
    \hline

    \end{longtable}
    \end{center}
\clearpage
\twocolumn

\subsubsection{Healthcare Provider}
In this section, popular healthcare companies in California such as Kaiser \cite{kaiserwebsite}, Blueshield \cite{blueshieldwebsite}, and Oscar \cite{oscarwebsite} were reviewed . All these websites do not have notice of collection when consumers visit. In addition, the Kaiser website - healthy.kaiserpermanente.org and Blueshield does not mention CCPA. Kaiser says they accept users' requests for personal information. However, they also reserve the right to deny these requests if the information is essential to the treatment. There is a toll-free number, but the website does not specify that number can exercise CCPA rights. Only Oscar company has a dedicated privacy page for CCPA, allowing users to exercise their data rights.   

Explaining the absence of CCPA on healthcare providers' websites, CCPA designs an exemption around the federal Health Insurance Portability and Accountability Act (HIPAA) \cite{hipaa}. HIPAA exempts a certain kind of protected health information (PHI). PHI is information relating to an individual's physical or mental health or condition or the provision of or payment for health care to an individual. There is a reasonable basis to believe it can be used to identify the individual. In addition, HIPAA exempts certain kinds of information and organizations that maintain patient information in the right way. Under HIPAA, a healthcare provider might be exempt as a whole; all of its non-healthcare information might qualify for CCPA's HIPAA exemption as long as healthcare providers have a proper way to protect "patient information."

The Organization's status under HIPAA and the purpose for collecting data will affect its eligibility to qualify for the CCPA's HIPAA exemption. For example,  a company that makes wearable devices capture people's personal information such as name, weight, birthday, and location. That company may not be covered under CCPA's HIPAA exemption. However, suppose a health care provider makes an app with the exact same functions but available only to patients to monitor their health condition for treatments. In that case, that Organization is covered under HIPAA. 

Now we can understand why many healthcare providers do not publish a CCPA privacy notice page on their websites. Instead, they only have a HIPAA notice page because it is qualified enough for a CCPA exemption in typical situations. In addition, California has amended CCPA to further harmonize with HIPAA to provide additional exemptions \cite{ccpamendment}. However, HIPAA is not a silver bullet to protect a company when incidents occur. Judges and assistant state attorneys may review a data incident to decide which portion of the data category is not PHI. In addition, healthcare providers tend to put information not exactly PHI under the broad category of "patient information." For instance, IP addresses, cookies, and marketing data of consumers who have not yet become patients may not fall under HIPAA. Therefore, the outcome in the CCPA criteria below is only supplemental because most of the healthcare providers in this research are exempted from CCPA. 

\clearpage
\onecolumn

\begin{center}
\begin{longtable}[c]{|p{\mycolwidth}|*{3}{p{\mycolwidth}|}}
\caption{CCPA criteria table for healthcare companies} \label{ccpa:long}\\
\hline 
\multicolumn{1}{|c|}{\textbf{}} & \multicolumn{1}{c|}{\textbf{Kaiser Permanent}} & \multicolumn{1}{c|}{\textbf{Blue Shield}} & \multicolumn{1}{c|}{\textbf{Oscar}}\\ \hline 
\endfirsthead

\multicolumn{3}{c}%
{{\mdseries \tablename\ \thetable{}: Continued from previous page}} \\
\hline 
\multicolumn{1}{|c|}{\textbf{}} & \multicolumn{1}{c|}{\textbf{Kaiser Permanent}} & \multicolumn{1}{c|}{\textbf{Blue Shield}} & \multicolumn{1}{c|}{\textbf{Oscar}}\\ \hline 
\endhead

\hline \multicolumn{3}{|r|}{{Continued on next page}} \\ \hline
\endfoot
\hline \hline
\endlastfoot
    \hline
    The company informs consumers what categories they have collected ?
    & Yes & Yes & Yes \\
    \hline
    Does the company inform consumers which specific pieces of personal information they have collected ? 
    & Yes & Yes & Yes \\
    \hline 
    The company verified the identity of consumers who requested to access or delete their personal information ? 
    & The website does not offer a way to access/delete information  & The website does not offer a way to access/delete information  & Yes \\
    \hline
    Does the company inform consumers before the point of collection ? 
    & No, as not being covered by CCPA
    & No, as not being covered by CCPA 
    & No notice of collection when visiting the website  \\
    \hline
    Does the company deliver information to consumers free of charge within 45 days ?
    & No, not being covered by CCPA   & No, not being covered by CCPA  & No, not being covered by CCPA  \\ 
    \hline
    The information the company has to deliver is portable and in a usable readable format that allows consumers to transmit the information to other entities without hindrance ?
    & No, not being covered by CCPA  & No, not being covered by CCPA   & No, not being covered by CCPA  \\ 
    \hline
    The company delete personal information when a consumer request it.
    & Yes, but the company reserve the right to reject the request because of not being covered by CCPA  & Yes, but the company reserve the right to reject the request because of not being covered by CCPA  & Yes  \\ 
    \hline
    The company has to create a process and to identify individuals responsible for consumers' opt-out and, therefore, not selling their data to third parties in response. 
    & No, not being covered by CCPA
    & No, not being covered by CCPA
    & No, not being covered by CCPA \\ 
    \hline
    By default, the company should not sell consumers' personal information when they're between 13 and 16 years old, but there has to be a process to allow them to opt-in. 
    & Yes  & Yes  & Yes   \\
    \hline
    The company has to provide consumers the right to equal services and prices. 
    & Yes & Yes & Yes \\
    \hline
    The company has to make available two or more designated methods for the consumer to request their information, including, at a minimum, a toll-free telephone number and website address (if the business maintains a website).
    & There are toll-free numbers and mailing addresses, but the company does not indicate that customers can use those to request their information, only for questions. There are toll-free numbers and mailing addresses, but the company does not indicate that customers can use those to request their information, only for questions & Yes & Yes \\
    \hline
    The company has to train and inform dedicated personnel to process new requests to exercise privacy rights properly.
    & Cannot verify   
    & Cannot verify   
    & Cannot verify  \\
    \hline
    The company has to disclose the consumer's rights to request the deletion of their personal information.
    & No & No & No \\
    \hline
    In case the company sells consumers' personal information have to inform your customers that their information may be sold and that they have the "right to opt-out" of the sale of their personal information.
    & No, the company says they do not sell data 
    & No, the company says they do not sell data. 
    & No, the company says they do not sell data  \\
    \hline
    If the company offers financial incentives for the collection, the sale, or the deletion of personal information, you need to disclose those financial incentives to your consumers.
    & No, the company says they do not sell data   
    & No, the company says they do not sell data   
    & No, the company says they do not sell data   \\
    \hline
    The homepage website has to include a link to inform consumers that they have the right to opt-out of their information sale.
    & No, not being covered by CCPA 
    & No, not being covered by CCPA 
    & No link on the home page, but there is a link in a Privacy notice. Users have to take one more step to the Privacy Notice link. 
 \\
    \hline
    The company has to disclose in its online privacy policy a description of consumer's rights and the categories of the consumer's personal information collected and sold in the preceding 12 months.
    & Yes, but Kaiser also says they have a right to deny a consumer's request if they think the requested information is essential for treatment purposes & Yes  & Yes \\
    \hline

    \end{longtable}
    \end{center}  
    
\clearpage
\twocolumn

\subsubsection{Academic Institution }
This section is about the evaluation of CCPA in an educational institution. We know that colleges and universities possess students' and staff's sensitive data that may be entitled to consumers' privacy rights. Therefore, we chose California State Polytechnic University, Pomona (CPP), to perform CCPA research on their website.
Generally, the CCPA is interpreted not to cover colleges and universities because they are often considered non-profit entities \cite{ccpauniversity}. For that reason, the CCPA notice is absent on CPP's website. However, CPP is proactive in protecting user data by having a well-written online privacy notice that overlaps many CCPA requirements \cite{cppprivacy}. Therefore, we rate the non-CCPA privacy notice that CPP currently posts on its website as a suitable substitution for CCPA privacy because of its comprehension. 

Following that, CPP states that they collect and process personal information such as the user's name, social security number, physical address, financial matter, and additional user-provided information. In the notice, CPP also explains how they use the information we collect. Some purposes include application, registration, service request processing, and fraudulent detection. In addition, CPP confirms that they do not collect personally identifiable information about users' online activities across third-party websites. Finally, CPP informs that they train their staff on procedures for managing personal information. Although CCPA does not cover CPP, the university still allows users to perform reviewing, updating and making changes to the personal information that the university maintains. Following that, we still use the criteria table to compare the privacy notice on their website with CCPA to see their perspective on handling users' data.  

\clearpage
\onecolumn

\begin{center}
\setlength{\mycolwidth}{\dimexpr \textwidth/2 - 2\tabcolsep}%
\begin{longtable}{|p\mycolwidth|p\mycolwidth|}
\caption{CCPA criteria table for academic institution } \\
\hline 
\multicolumn{1}{|c|}{\textbf{}} & \multicolumn{1}{c|}{\textbf{Cal Poly Pomona }}\\ \hline 
\endfirsthead

\multicolumn{1}{c}%
{{\mdseries \tablename\ \thetable{}: Continued from previous page}} \\
\hline 
\multicolumn{1}{|c|}{\textbf{}} & \multicolumn{1}{c|}{\textbf{Cal Poly Pomona}}\\ \hline 
\endhead

\hline \multicolumn{1}{|r|}{{Continued on next page}} \\ \hline
\endfoot
\hline \hline
\endlastfoot
    \hline
    The company informs consumers what categories they have collected ?
    & Yes\\
    \hline
    Does the company inform consumers which specific pieces of personal information they have collected ? 
    & Yes \\
    \hline 
    The company verified consumers' identity who requested to access or delete their personal information ? 
    & Yes\\
    \hline
    Does the company inform consumers before the point of collection ? 
    & No notice of collection on the homepage (CCPA does not cover the Organization) \\
    \hline
    Does the company deliver information to consumers free of charge within 45 days ?
    & Yes\\
    \hline
    The information the company has to deliver is portable and in a usable readable format that allows consumers to transmit the information to other entities without hindrance ?
    & Yes\\
    \hline
    The company delete personal information when a consumer request it.
    & Yes\\
    \hline
    The company has to create a process and to identify individuals responsible for consumers to opt-out and, therefore, not selling their data to third parties in response.
    & Yes, although the Organization does not sell user's data \\
    \hline
    By default, the company should not sell consumers' personal information when they are between 13 and 16 years old, but there has to be a process to allow them to opt-in.
    & Yes, although the Organization does not sell user's data \\
    \hline
    The company has to provide consumers the right to equal services and prices.
    & Yes \\
    \hline
    The company has to make available two or more designated methods for the consumer to request their information, including, at a minimum, a toll-free telephone number and website address (if the business maintains a website).
    & Yes \\
    \hline
    The company has to train and inform dedicated personnel to process new requests to exercise privacy rights properly.
    & Yes, the Organization commits to training staff on protecting user's data\\
    
    \hline
    The company has to disclose the consumer's rights to request the deletion of their personal information.
    & Yes \\
    \hline
    Suppose the company sells consumers' personal information. In that case, it has to inform your customers that their information may be sold and that they have the "right to opt-out" of the sale of their personal information.
    & Organization says they do not sell data but allow users to have their information discarded\\
    \hline
    If the company offers financial incentives for the collection, the sale, or the deletion of personal information, you need to disclose those financial incentives to your consumers.
    & Organization says they do not sell data \\
    \hline
    The homepage website has to include a link to inform consumers that they have the right to opt-out of their information sale.
    & Yes, but Organization says they do not sell data  \\
    \hline
    In its online privacy policy, the company has to disclose a description of consumers' rights and the categories of consumers' personal information collected and/or sold in the preceding 12 months.
    & Yes  \\
    \hline

    \end{longtable}
    \end{center}
\clearpage
\twocolumn
\subsection{Evaluate Effortlessness of Exercising Consumer's CCPA Rights}
\subsubsection{Social Media Websites}
User can access their Facebook information through the Facebook setting. We can also request a copy of our data in HTML or JSON format through the portal. 
The process takes approximately 10 minutes; then, a link is provided to download our data in zip format. Extracting that zip file gives us a list of Facebook's data collection presented as folders. We conclude that requesting data that can be performed using the Facebook portal is pretty straightforward. A screenshot of the data Facebook gave us is as follows: 
\begin{figure}[h]
    \centering
    \includegraphics[width=0.3\paperwidth]{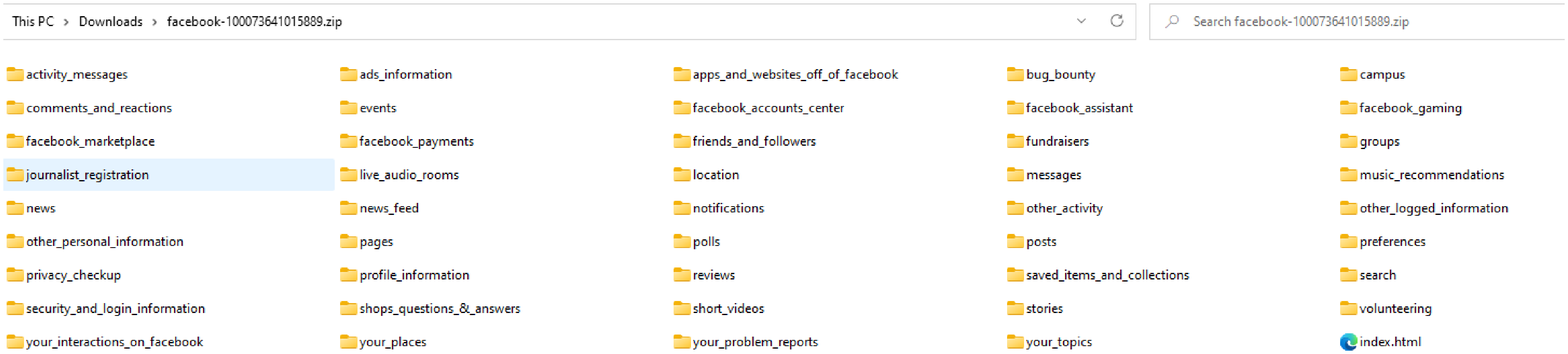}
    \caption{JSON data provided by Facebook}
    \label{fig:my_label}
\end{figure}

However, recent research shows that Facebook still keeps collecting people's data and building their profiles even when their accounts are deactivated \cite{deactivedaccountdatacollection}. Deactivation can only make their profile disappear from friends but not Facebook. Facebook handles deactivated accounts just like an active account. Hence, they are still able to append data to accounts. During deletion, Facebook does not delete the user's account immediately. Instead, the deletion would be scheduled. Facebook requires 30 days in which people cannot log back into the account before Facebook begins deleting their data. Because of that, there are many ways that a user may accidentally log into their account again through SSO login that other websites utilize Facebook credentials to run their services \cite{singlesignon}. As soon as users log in to their accounts, the deletion process will be canceled.

\subsubsection{Bank Website }
Although Bank Of America implements CCPA, when we requested to access our data, they responded with a notice "CCPA does not apply for consumer accounts used for personal, family, or household purposes." In addition, BOA did not give us a detailed report of our collected data. Instead,  BOA only provided us with the categories of personal information they collected on our account. 

We made another request to access our data to another bank, Chase. Two days later, they could send us a data collection report on our account. Initially, the differences between the response of Bank Of America and Chase raised a question about if financial institutions have a right to reject our data access request based on the type of our account. Being explained earlier, we understood that BoA might apply the exemption provided by the federal Gramm-Leach-Bliley Act\cite{financialinstitutionandccpa}. However, there are still liabilities that BoA may face in case of a data leak. 
\begin{figure}[h]
    \centering
    \includegraphics[width=0.4\paperwidth]{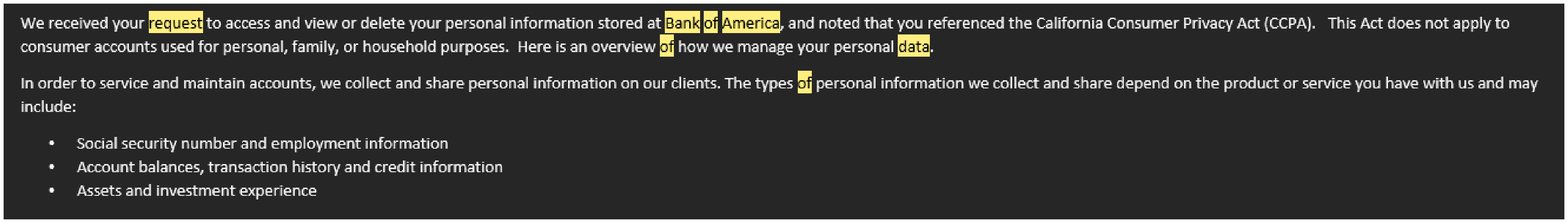}
    \caption{CCPA response from Bank of America}
    \label{fig:my_label}
\end{figure}

\section{Research Goal}
\label{sec:research_goal}
To verify CCPA on a website, consumers have no other way except go to the website they want to check and then read through pages of the privacy notice to figure out which CCPA criteria the company satisfies. That process is time-consuming, so we believe that technology steps in to provide consumers with a more effective method to verify CCPA, which is also the goal of our research. We strive to develop a feature which is called CCPA compliance verifier. The application can provide online consumers with an effortless approach to verifying CCPA compliance on websites. The application should function with any website to crawl data regardless of web technology. Based on the verifier's result, the application can assist website admins in adjusting their website to comply with CCPA. The feature requires the application to generate accurate suggestions to fix the website's current issues to reach the satisfactory status of CCPA. 

This project's final goal is to construct an intuitive user interface and integrate it with the back-end application. After that, we'll deploy both the back-end and front-end for public access. Finally, our application will eliminate the nuisances that other online tools demand users to register an account or provide unneeded personal information to receive a CCPA compliance report. To achieve that goal, we will apply a modern web API design pattern to maximize the reusability of app components and reduce clusters between application versions. 

By comprehending the importance of user privacy and the effectiveness of CCPA, this project will concentrate on creating a robust tool to inform and aid online users in exercising their data rights. Furthermore, we confidently state that our application can reduce the significant amount of time consumers need to verify a website's CCPA compliance. Accordingly, companies would have more inducement to comply with CCPA when consumers have efficient tools to exercise their CCPA rights. In short, this project is a win-win solution for both consumers and companies by enhancing data privacy on the consumer's side and help preventing legal issues on the organization's side.

\section{\centering Methodology}
\label{sec:methodology}
There are three features to be completed in this project. The first is building a web application to verify CCPA on a website. The application will present essential information for consumers to understand whom CCPA applies to and what are consumers' rights.

The next feature is integrating a web crawler framework such as Java Selenium to take a website URL and crawl data from that website to look for CCPA implementation on a website. For each CCPA criteria, we will apply natural language processing to detect CCPA criteria keywords. As soon as the web crawler gathers enough information about CCPA that can help with its decision, the application will inform users if the website satisfies that CCPA criteria. The application may also display the website's screenshot, which is being tested for extra validation. 

The last feature is to have a CCPA compliance helper tool that can recommend changes to the website owner to comply with CCPA fully. According to the California attorney general's office, violations of the CCPA can have civil penalties of \$2,500 for each violation or \$7,500 for each intentional violation \cite{CCPAEnforcing}. Without a doubt, companies should adapt to CCPA quickly to avoid such penalties. Based on the CCPA checking feature, the tool will generate the list of modifications the website needs to complete to fulfill CCPA criteria. Then, the website admin can refer to the recommendation to make necessary modifications.  

\subsection{ Application Business Requirement}
\label{sec:business_requirement}
\subsubsection{Scope}
\begin{enumerate}
    \item Project Scope
    \begin{itemize}
        \item Initial system
        \begin{itemize}
            \item Will be a single page application web application
        \end{itemize}
        \item Initial Browser Support
        \begin{itemize}
            \item Will support latest version of Chrome, Firefox or Edge
        \end{itemize}
        \item Initial Data
        \begin{itemize}
            \item Website URL : The application will accept website URL as input string to perform CCPA verifying function  
        \end{itemize}
        \item Permission for collecting user data 
        \begin{itemize}
            \item To monitor the application performance, the application may record the following data from consumers :
            \begin{itemize}
                \item IP address
                \item Fail and Successful operations
                \item Website URL being checked
                \item Responding time
            \end{itemize}
        \end{itemize}
    \item In Scope Functionality
    \begin{itemize}
        \item Presenting of CCPA Information
        \item CCPA Compliance Verifier 
        \item Generate Instruction To Exercise CCPA Rights
        \item Generate Instruction To File CCPA Compliant 
    \end{itemize}
    \item Out of Scope Functionality
        \begin{itemize}
            \item Develop an Android version of the application
            \item Allow consumers to perform mass checking on multiple websites at same time. 
        \end{itemize}
    \end{itemize}
    
    \item Constraints
    \begin{itemize}
        \item Time
        \begin{itemize}
            \item Our final deadline is May 15
            
        \end{itemize}
        \item Resources
        \begin{itemize}
            \item Our team is compiled of one member
        \end{itemize}
    \end{itemize}
\end{enumerate}

\subsubsection{Project Function and Non-functional Requirements}

\clearpage
\onecolumn

\begin{center}
\setlength{\mycolwidth}{\dimexpr \textwidth/2 - 2\tabcolsep}%

\begin{longtable}{|p\mycolwidth|p\mycolwidth|}
\caption{Application's function and non-functional requirements } \label{ccpa:long}\\
\hline 
\multicolumn{1}{|c|}{\textbf{Functional Requirements}} & \multicolumn{1}{c|}{\textbf{Non Functional Requirements }}\\ \hline 
\endfirsthead

\multicolumn{2}{c}%
{{\mdseries \tablename\ \thetable{} -- Continued from previous page}} \\
\hline 
\multicolumn{1}{|c|}{\textbf{Functional Requirements}} & \multicolumn{1}{c|}{\textbf{Non Functional Requirements }}\\ \hline 
\endhead

\hline \multicolumn{2}{|r|}{{Continued on next page}} \\ \hline
\endfoot
\hline \hline
\endlastfoot
    \hline
    One web page is used to house the application
    
    & \\
    \hline
    The page changes to display a new section based on user interactions.
    
    & Page loading of a new section of the web app should take less than 3-second \\
    \hline
    The web application is deployed on a cloud server to allow anyone to access the page online.
    
    & Cloud service should not take down our application for daily maintenance any longer than thirty minutes\\
    
    \hline
    Cloud service should be resistant to most common attacks.
    
    & \\

    \hline
    The application should have a modular design to allow any part to be modified without needlessly affecting any other part.
    
    & \\
    
    \hline
    The part should be built to allow a unit test to be created for each. 
    
    & \\
\end{longtable}
\end{center}

\clearpage
\twocolumn
\subsubsection{Business Requirements (Core Components)}
\paragraph{}{Security}
\begin{itemize}
    \item \textbf{Background}: Current version of the application does not save any user data. Therefore, there is no risk of leaking data at this moment. However, the application still utilizes HTTPS protocol \cite{HTTPS} to encrypt the data transferred between the backend server and the front-end application. 
    In addition, the backend application should only receive requests from our domain at ccpaverifier.com and reject requests from other domains to protect data integrity and the backend server itself. Cross-Origin Resource Sharing (CORS) \cite{CORS} technique will be used for that purpose.
    \item \textbf{Actor}: Consumers.
    \item \textbf{Precondition}: The program is released, and consumers start to input the website URL into a typing box.
    \item \textbf{Post condition}: The website URL is encrypted using SSL and sent over port 443. Results, including screenshots, would be transferred back to the source using the same method.
    \item \textbf{User Stories:} "As a consumer, I want the data that I send and receive to and from the backend server to be encrypted  and must not be seen or modified by anyone"
    \item \textbf{Business Rules:}
    \begin{itemize}
        \item Encrypt transferred data.
        \item The application should be able to handle penetration testings using the OWASP framework or similar frameworks.
        \item Man in Middle cannot read the data transferred between source and destination. The result cannot be altered.
        
    \end{itemize}
    \item \textbf{Pass Criteria:}
    \begin{itemize}
        \item Transferred data can only be  encrypted and decrypted at the front-end app and back-end server.
        \item Website can survive from DDOS attack or other common web attacks.
    \end{itemize}
    
    \item \textbf{Fail Criteria:}
    \begin{itemize}
        \item Data is not encrypted before being transferred over the internet. Attacker can read the transferred data.
        \item The web server cannot resist common wed attacks. 
    \end{itemize}

\end{itemize}

\paragraph{Feature to present CCPA Content}
\begin{itemize}
    \item \textbf{Background}: This feature include two components. First, the website will display essential information about CCPA. Second, if consumers want to learn more about CCPA, they can click on button which contain a link on the website to redirect themselves to the official page of CCPA  under  Office of Attorney General website. 
    \item \textbf{Actor}: Consumers.
    \item \textbf{Precondition}: The website is up and running. Consumer can access the homepage.
    \item \textbf{Postcondition}: Consumers can read information about CCPA or visit the official CCPA website
    \item \textbf{User Stories:} "As a consumer, I want to learn about CCPA to understand which organizations it applies to and my rights"
    \item \textbf{Business Rules:}
    \begin{itemize}
        \item Our website must display information about CCPA in a readable format
        \item Our website must redirect consumers to official CCPA  page immediately when consumers click on the button which contain a link.
        \begin{center}
\setlength{\mycolwidth}{\dimexpr \textwidth/2 - 2\tabcolsep}%
\clearpage
\onecolumn
\begin{longtable}{|p\mycolwidth|p\mycolwidth|}
\caption{Function and non-functional requirements of CCPA content presentation feature } \label{ccpa:long}\\
\hline 
\multicolumn{1}{|c|}{\textbf{Functional Requirements}} & \multicolumn{1}{c|}{\textbf{Non Functional Requirements }}\\ \hline 
\endfirsthead

\multicolumn{2}{c}%
    {{\bfseries \tablename\ \thetable{} -- continued from previous page}} \\
\hline 
\multicolumn{1}{|c|}{\textbf{Functional Requirements}} & \multicolumn{1}{c|}{\textbf{Non Functional Requirements }}\\ \hline 
\endhead

\hline \multicolumn{2}{|r|}{{Continued on next page}} \\ \hline
\endfoot
\hline \hline
\endlastfoot
    \hline
   After visiting the home page, the user can scroll down to read essential information about CCPA. 
    
    & CCPA information is displayed in an easy-to-read font and appropriate font size \\
    \hline
    To access the official CCPA website to learn more about CCPA, consumers need to click on the  "Information about CCPA" card. 
    
    & The arrow button will appear as soon as consumers hover the mouse on the content holder component \\

\end{longtable}

\clearpage
\twocolumn
\end{center}
    \end{itemize}
    \item \textbf{Pass Criteria:}
    \begin{itemize}
        \item Summarized information about CCPA is presented correctly and clearly
        \item Consumers are redirected to official CCPA  website  upon request
    \end{itemize}
    
    \item \textbf{Fail Criteria:}
    \begin{itemize}
        \item The website failed to display CCPA information.
        \item the website failed to redirect consumers to official CCPA site .
    \end{itemize}

\end{itemize}

\begin{figure}[H]
    \centering
    \includegraphics[width=40mm]{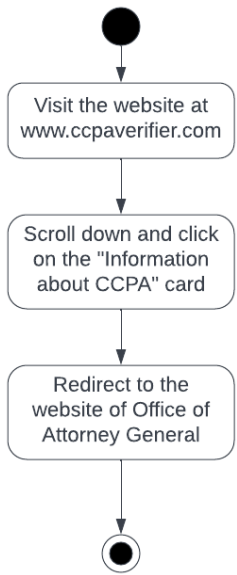}
    \caption{Presenting of CCPA information flow}
    \label{fig:my_label}
\end{figure}
\paragraph{CCPA Compliance Verifier}
\begin{itemize}
    \item \textbf{Background}: This feature allows consumers to choose a website from a list or enter their own URL to verify CCPA compliance of the website. Then the application will utilize  web crawling Selenium to check if it can find information that match the list of CCPA requirements. 
    \item \textbf{Actor}: Consumers.
    \item \textbf{Precondition}: The website available to access. Consumers can click on the accessing feature button to use the CCPA verifier. 
    \item \textbf{Postcondition}: CCPA verifier will provide consumers result after finishing CCPA verifying function. 
    \item \textbf{User Stories:} "As a consumer, I want to pick a website from the list to verify if that website complies with CCPA", " As a consumer, I want to enter a website URL to check if that website complies with CCPA" 
    \item \textbf{Business Rules:}
    \begin{itemize}
        \item The application must present a result to consumers in no longer than one minute.
        \item The application must show an error if back-end server is not working or consumers submit wrong URL.
        \item The application must generate correct result.
        \item The result the application presents to user must be freshly generated at the time of submission. It means the result must be original, not be stored prior to the current session then retrieved.  
        \item If the application decides that the tested website complies with CCPA, our application must present a link to the CCPA page. Then consumers will able to use that link to exercise their CCPA rights. 
        \begin{center}
\setlength{\mycolwidth}{\dimexpr \textwidth/2 - 2\tabcolsep}%

\onecolumn
\begin{longtable}{|p\mycolwidth|p\mycolwidth|}
\caption{Function and non-functional requirements CCPA compliance verifier } \label{ccpa:long}\\
\hline 
\multicolumn{1}{|c|}{\textbf{Functional Requirements}} & \multicolumn{1}{c|}{\textbf{Non Functional Requirements }}\\ \hline 
\endfirsthead

\multicolumn{2}{c}%
{{\mdseries \tablename\ \thetable{} -- Continued From Previous Page}} \\
\hline 
\multicolumn{1}{|c|}{\textbf{Functional Requirements}} & \multicolumn{1}{c|}{\textbf{Non Functional Requirements }}\\ \hline 
\endhead

\hline \multicolumn{2}{|r|}{{Continued On Next Page}}\\
\endfoot
\hline \hline
\endlastfoot
    \hline
    Consumers can either choose a website from the list or enter a website URL manually. The front-end app will send a request to the backend server to perform CCPA verifying tasks. 
    
    &  User can enter website URL with our without prefix 'http://www.' The application can detect the missing parts and add them automatically. \\
    
    \hline
    If CCPA is found on the tested website, the application shows a link for consumers to exercise  CCPA rights, such as how to request data access, data deletion, or opt-out of sale. 
    & The front end UI shows the "click here" button that redirects the user to the CCPA page of the tested website. \\
     
    \hline
    Suppose CCPA is not detected on the website being tested. In that case, the application will let consumers download a result of a screenshot and then instruct them to file a complaint to the Office of the Attorney General.
    &
    The front-end interface shows the "Show Instruction" button that displays a dialog box instructing consumers how to file a complaint. \\
    
    \hline
    The function is asynchronous, so consumers can still interact with other UI components during the ongoing verifying process. 
    
    & While waiting for a response, the application will display spinning wheels to let users know the process is still ongoing  \\

\end{longtable}

\end{center}
    \end{itemize}
    \clearpage
\twocolumn
    \textbf{Pass Criteria:}
    
        - The application can present a result after finishing verifying CCPA on a website \\
        - The application can present a screenshot of the tested website \\
        - The application can present a link for consumers to exercise CCPA rights if the website complies with CCPA\\
        - The application can show instructions for filing a complaint if the tested website does not comply with CCPA.\\

    \textbf{Fail Criteria:}
  
       - The application displays an error because the backend server does not respond.\\
       - It takes longer than the expected amount of time for the backend server to respond to the request.\\
       - The web application gives incorrect results compared with a result acquired manually. \\

\end{itemize}

\begin{figure}[H]
    \centering
    \includegraphics[width=0.3\textwidth]{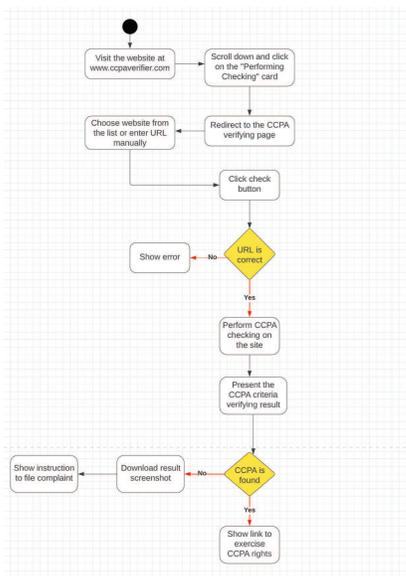}
    \caption{CCPA criteria verifying flow}
    \label{fig:my_label}
\end{figure}

\section{Design}
\label{sec:design}
\subsection{Dependency Diagram}
This section presents a dependency diagram of our web application. The dependency diagram explains how different application features rely on each other. For example, feature A depends on feature B. When feature B is unavailable, feature A will not work as specified in the business rule. According to the dependency diagram, a notice of collection checking, right to know to check, right to delete checking, and right to opt-out of sale matching feature rely on the CCPA notice checking feature, which relies on privacy notice checking. Then the features which offer assistance to exercise CCPA rights, filing CCPA complaints, and capturing a screenshot of the result all depend on the CCPA rights checking mentioned earlier.
\begin{figure}[H]
    \centering
    \includegraphics[width=0.4\textwidth]{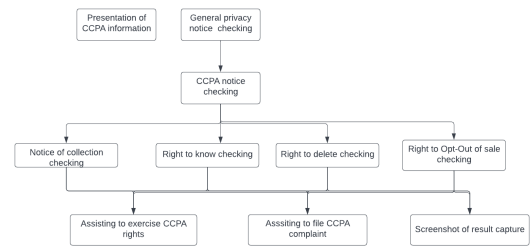}
    \caption{Dependency diagram of features}
    \label{fig:my_label}
\end{figure}

\subsection{High Level Flow Diagram}
\subsubsection{Introduction}
This section explains the flow of the data with high-level architecture. We have divided our web application into different layers, and each of them handles the data flow in our application. Our application consists of six layers: UI, Router, Data Handler, and Service. Each of these layers takes a request and return a variety of response depending on the given request. A layer communicates directly with its surrounded layers only to lower the complexity level of our web application. We will explain the behavior of each layer in detail later. In addition to those layers, some layers affect multiple layers directly. There is security and error handling. The technology that the backend uses to communicate with the front-end is REST \cite{restapi}

\begin{figure}[H]
    \centering
    \includegraphics[width=0.4\textwidth]{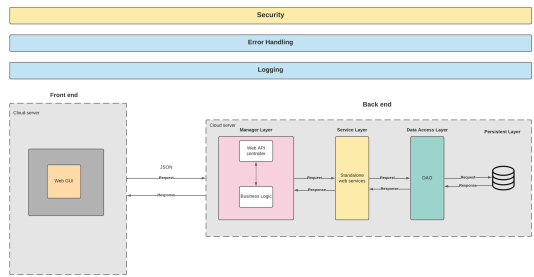}
    \caption{High-level architecture diagram}
    \label{fig:my_label}
\end{figure}

\subsubsection{User Interface Layer (UI)}
The front-end of our web application is what we use to present visual content to consumers. This layer directly interacts with the user and accesses the server to fulfill the valid request. It forms a request, sends it to the server, and renders any response received from the server.

\subsubsection{Service Layer}
The service layer consists of standalone services that meet the consumers' requests. It takes input from the manager layer and processes it before returning the result. It may use the lower level to store and access data if necessary. Standalone services should be as general as possible so we can reuse them in other parts of the application.

\subsubsection{Security}
Considering security in high-level design is critical to protect our servers and ensure the integrity of the result generated by the application. Although the application does not save sensitive data in this version, the data generated by the application must be encrypted before transferring to avoid being altered. We enforce using HTTPS protocol between the front-end and backend servers to do that. HTTPS protocol uses public-private key infrastructure to encrypt and decrypt information when sending and receiving data. \\
In addition, the cloud servers need to be secured to avoid being attacked. We only allow necessary ports opened, and the backend server only accepts requests from our front-end domain using a cross-origin resources sharing (CORS) policy. To make the server even more secured, we think about allowing only US IP addresses to access our website. Then it may help put the website out of the radar of hackers outside the US.

\subsubsection{Error Handling Rules}
\begin{itemize}
    \item Business Logic : 
    \begin{itemize}
        \item When consumers attempt to connect to the server when it is offline, the application will show an error letting the consumers know the backend server is not functioning. 
        \item When consumers submit an invalid website address, the application will return a not found or bad input error and then ask consumers to submit again. 
    \end{itemize}
    
    \item UI
    \begin{itemize}
        \item When consumers y to input invalid data, the application will raise an exception as a result of invalid behavior
    \end{itemize}
    
    \item UI and Business Logic 
    An exception will be raised when the users try to act on application-specific business rules. Custom exceptions appropriate to a situation will be raised to produce a descriptive error message to users and administrators
    \end{itemize}

\subsubsection{Network Architecture Diagram}
In this network architecture design, we provide a high-level network architecture diagram. It explains the overall process of data flow in our networking system. The chart consists of the following network components:
Client
DNS server
Front-end firewall
Front-end web server
Backend firewall
Back-end web server
Generally, we deploy our backend and front-end apps on two different virtual machines and let them communicate over the internet. Separating the two servers, it will make the system more extensible. In the future, we can develop an Android or iOS application that will talk to the same backend app. In addition, when doing maintenance, we can work on either application without disturbing the other application. 
\begin{figure}[H]
    \centering
    \includegraphics[width=0.4\textwidth]{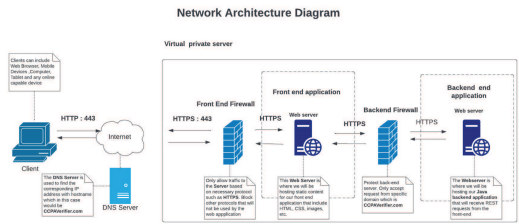}
    \caption{Network architecture diagram}
    \label{fig:my_label}
\end{figure}

\subsection{Low-Level Designs}
This section includes diagrams showing the data flow through the feature and its error handling charts. These diagrams explain the features we are currently building and the mistakes in certain areas.
\subsubsection{CCPA Verifier}
This design will show the data flow of the CCPA Verifying feature. Two scenarios may happen when executing the CCPA verifier. In the first scenario, if a CCPA privacy notice is found on the tested website, the application will perform additional functions such as checking for the right to access, delete, and opt-out. Then the application will return a link to the CCPA page of the website, where consumers can exercise their CCPA rights. In the second scenario, if the application cannot find any information related to CCPA on the tested website, it will allow consumers to download a screenshot of a website and then instruct consumers to file a complaint to the Office of Attorney General. 

\clearpage
\onecolumn
\begin{figure}[H]
    \centering
    \includegraphics[width=1\textwidth]{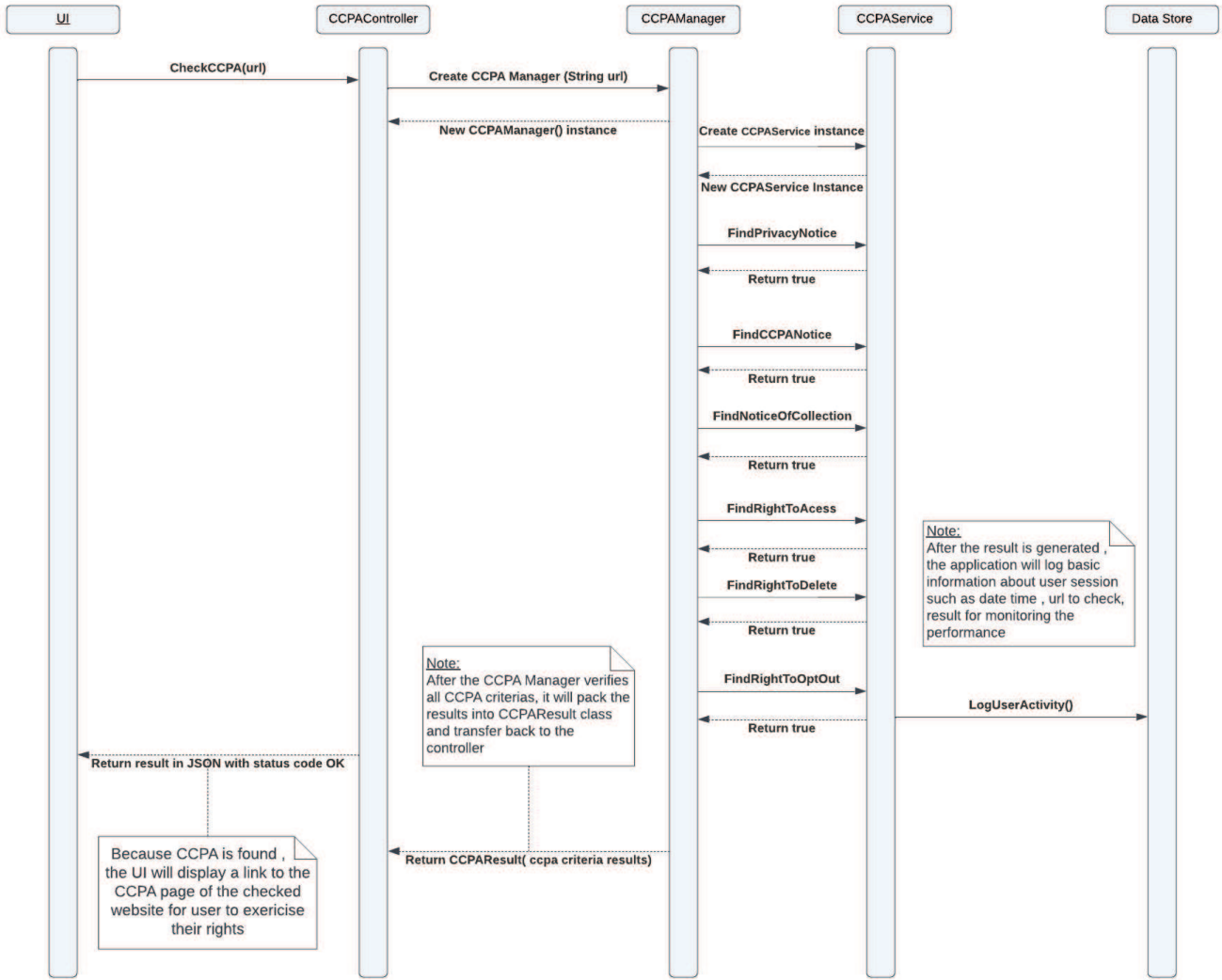}
    \caption{CCPA verifier sequence diagram - CCPA found }
    \label{fig:my_label}
\end{figure}

\begin{figure}[H]
    \centering
    \includegraphics[width=0.7\paperwidth]{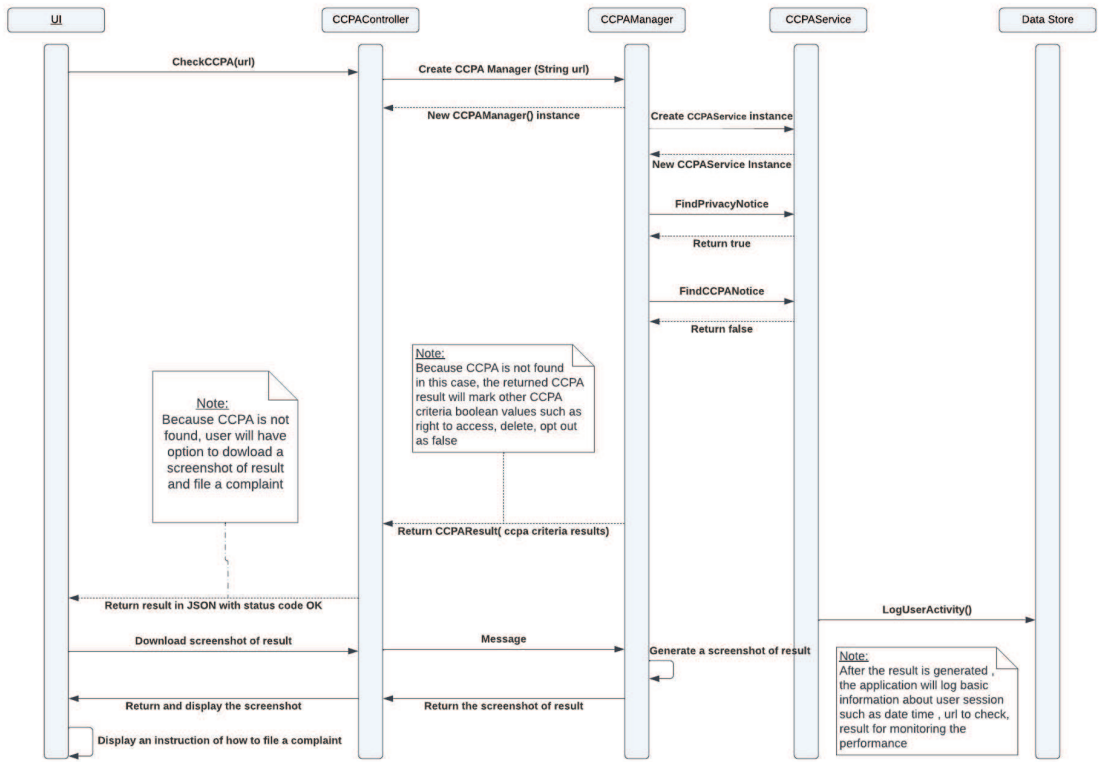}
    \caption{CCPA verifier sequence diagram - CCPA not found }
    \label{fig:my_label}
\end{figure}

\clearpage
\twocolumn

\section{ Project Implementation - CCPA Verifier 1.0 }
\label{sec:implemetation}
\subsection{Backend application}

The backend server is written in Java and follows using a micro-services pattern. The application is constructed by four major java packages: Main application file, controller package, manager package, model package, and service package. The below figure describes how we structure our application.

\begin{figure} [h]
    \centering
    \includegraphics[width=0.4\paperwidth]{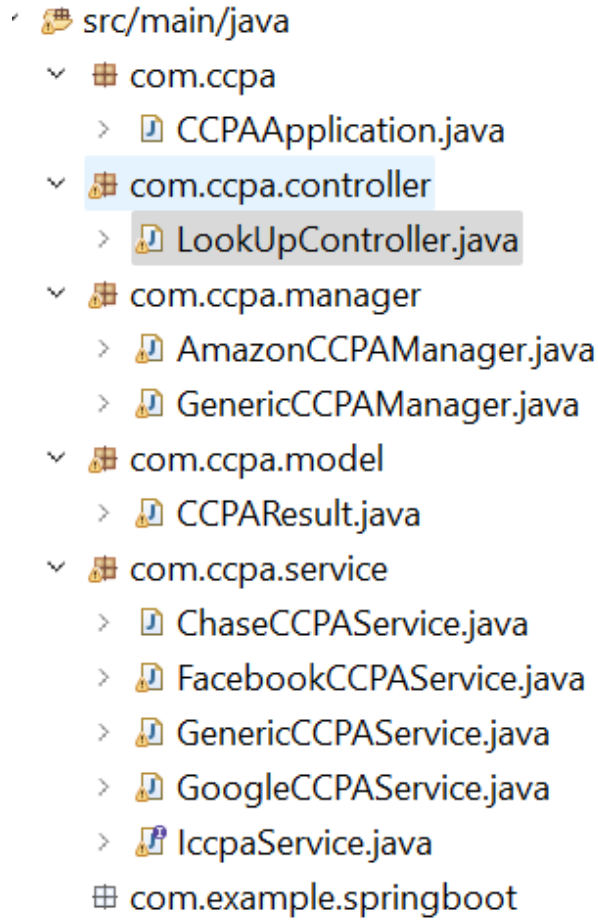}
    \caption{Application structure}
    \label{fig:my_label}
\end{figure}

The main application file is the start-up file that the application uses to execute the first functions. The service package contains standalone and reusable functionality that the app uses to check CCPA criteria on the website. For example, the generic CCPA service class has functions to look for notice of collection, CCPA link, right to access, right to delete, and opt-out. For more complicated websites which require a hybrid data crawling solution (manual direct the application which pages to crawl data first), we can extend the generic CCPA service to create a website-specific version of CCPA service. For example, we can create a CCPA service named ChaseCPPAService that specifically contains crawling functions working with the Chase website. To write an extended function is simple; we only need to input a URL that points to the site's CCPA privacy page directly to skip the automatic CCPA searching process that the generic CCPA function does. That approach would ensure the application can always reach the CCPA privacy page even if the website has a complex structure. After that, the rest of the process can still be automated by the generic CCPA module.
We usually do not need to override the CCPA criteria look-up functions in most cases. Currently, our application service package contains the following classes:  GenericCCPAService class, GoogleCCPAService, FacebookCCPAService, and ChaseService. We will add more classes to the application as the application needs to check more websites. We believe this project would benefit our community in the long term, so we aim to maximize the reusability and make the application more maintainable. 

Next, the manager package combines standalone services to satisfy application business requirements. For example, when consumers perform CCPA checking on a website, the application needs to execute a set of tasks that include looking for a privacy notice first, then CCPA notice, then right to access, right to delete, and right to opt-out, etc. Therefore, we use the manager class to schedule these functions' execution and then generate a complete result that includes data fields to indicate if the privacy notice page exists, the CCPA page exists, and CCPA rights exist. The result is packed in a model class called CCPAResult. After the result is generated, the manager classes return it to the controller getting the data ready to transfer to a front-end application. 

The last package in our application is the controller package. It is the gateway that the backend application receives web requests from a front-end application. The data is served under JSON format. To request data, the front-end application will send a REST request to the endpoint with the website URL as the request parameter. For performance, every CCPA criteria checking would be performed using a single REST request, so we only have two API endpoints now. One is used for the CCPA checking, and the other one is used to retrieve browser screenshots for the front end. 
\subsection{Front End Application}
The front-end application allows consumers to interact with the backend app using a graphical user interface. The front-end app is deployed on a cloud webserver. The website is developed as a single web page application with material design to give consumers an intuitive experience. The website is hosted at this URL: www.ccpaverifier.com. The homepage will present brief CCPA information to advertise the privacy law. The website structure consists of a landing page, a feature page for CCPA checking, a team information page, and a contact page.

\subsubsection{Landing Page}
The landing page is the first location consumers will experience when visiting the website. The landing page also includes other nested pages because of the nature of a single-page application (SPA). Therefore, that beauty is to allow consumers to scroll down to discover more website content without being redirected to other pages. In addition, SPA will enable consumers to take less time to understand the website structure. Only the feature page on which consumers perform CCPA checking on other websites is located on another page.

\begin{figure}[H]
    \centering
    \includegraphics[width=0.4\textwidth]{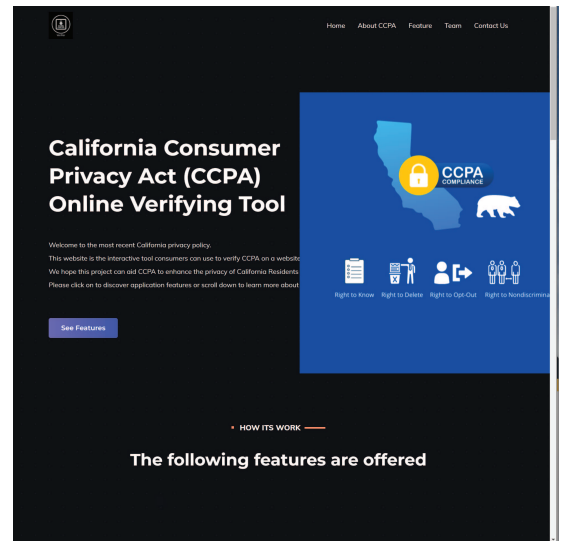}
    \caption{Landing page}
    \label{fig:my_label}
\end{figure}
\subsubsection{Feature Page}

This section introduces the features that our application offers. The first feature redirects consumers to the full CCPA document. The second feature allows consumers to select a website from a given list and then perform the CCPA verification. The last feature will enable consumers to enter the URL of any website. Then, the backend server will execute a generic CCPA verifying function to try its best to retrieve a result. If the application can generate the result, it will render the result on the website. Otherwise, the front-end application would present an error to notify consumers that the verifying process is failed and ask them to try another URL.

\begin{figure}[H]
    \centering
    \includegraphics[width=0.4\textwidth]{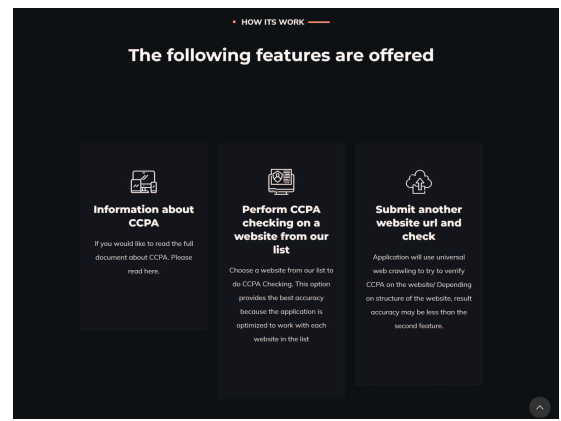}
    \caption{How it works page}
    \label{fig:my_label}
\end{figure}

\subsubsection{How it Work Page}
This page consists of a diagram that illustrates the business logic of the CCPA Verifier. Consumers who are not tech-savvy may question how the application works. Therefore, this page will provide a high-level, end-to-end flow of the feature.

\begin{figure}[H]
    \centering
    \includegraphics[width=0.4\textwidth]{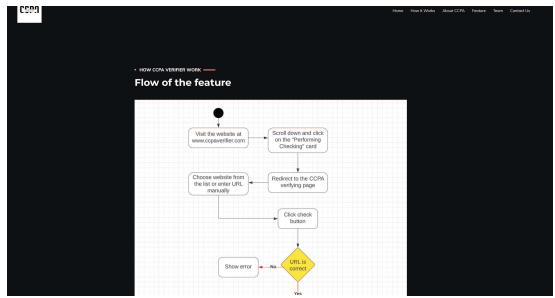}
    \caption{CCPA feature flow page}
    \label{fig:my_label}
\end{figure}

\subsubsection{CCPA Information Page}
This section presents essential information about CCPA, such as subjects that CCPA applies to, consumers' rights, and penalties that organizations may receive if they do not implement CCPA. This module aims to provide consumers who visit the site a quick preview of CCPA. If consumers want to know more about CCPA, they can follow the link on our site to read a full CCPA document.

\begin{figure}[H]
    \centering
    \includegraphics[width=0.4\textwidth]{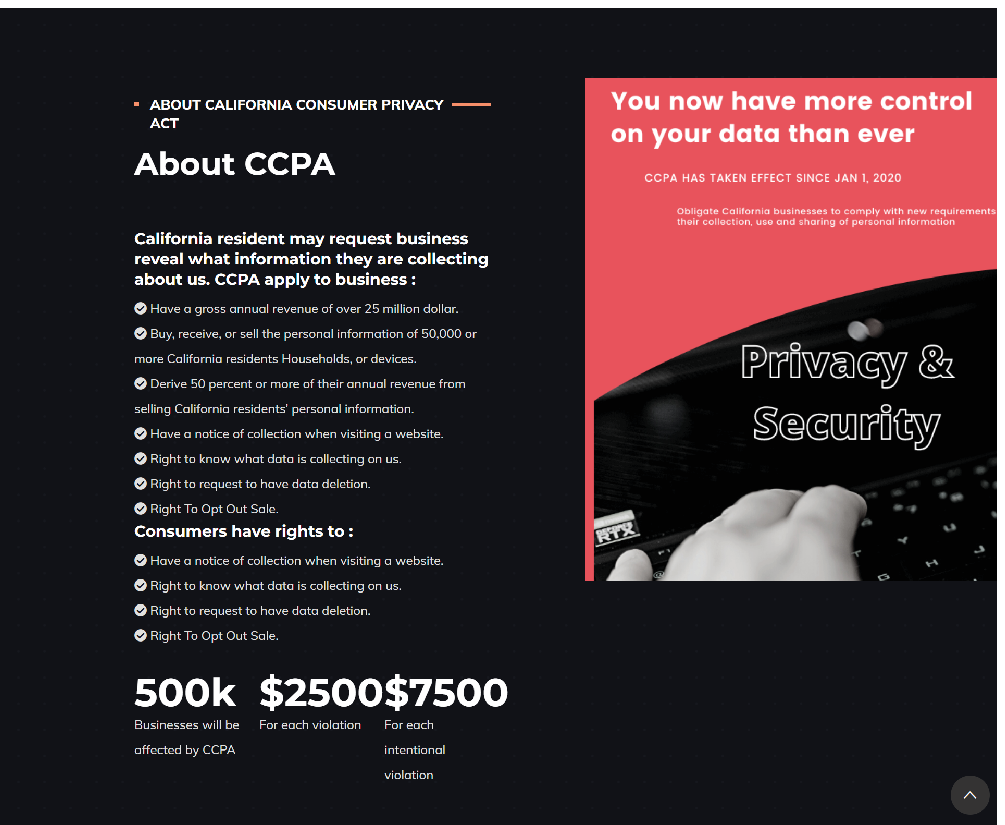}
    \caption{CCPA information page}
    \label{fig:my_label}
\end{figure}
\subsubsection{Contact Us Page}
Our application is still in developmental phase so we would like to receive feed backs to improve design as well as functionalities. Therefore, we include an contact form on our website and let consumers submit questions,feed-backs or report error. Based on the response, we will tweak and improve our application overtime.

\begin{figure}[H]
    \centering
    \includegraphics[width=0.4\textwidth]{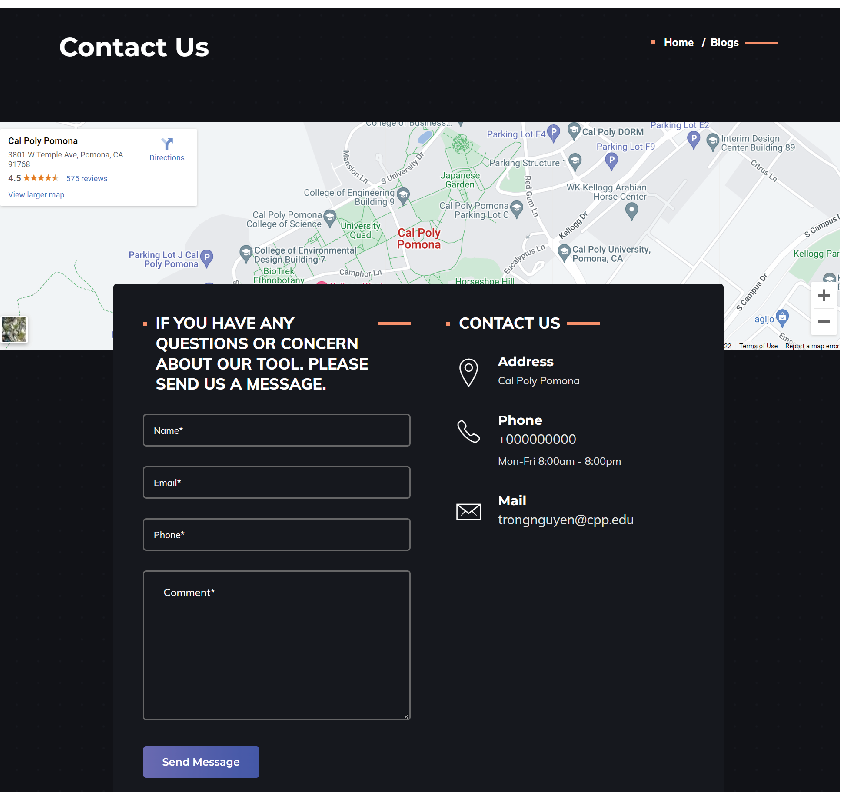}
    \caption{Contact us page}
    \label{fig:my_label}
\end{figure}

\subsubsection{Privacy Policy Page}
This page will inform consumers about our data practice, including what data we collect and how we use these data. Because our application crawls data from other websites or collects consumers' contact information, we must have a privacy policy to state clearly our purpose over that action. Other important information included in the policy is about using cookies,  declaration of not tracking consumers, and not sharing consumers' data. 
\begin{figure}[H]
    \centering
    \includegraphics[width=0.4\textwidth]{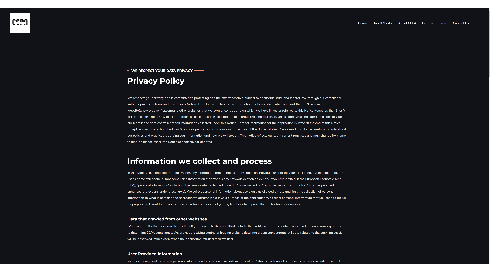}
    \caption{Privacy policy page}
    \label{fig:my_label}
\end{figure}

\subsubsection{Technologies used}
\label{subsec:Technologies}
\paragraph{Java Selenium}
\label{subsubsec:Selenium}
Selenium is a framework to make web automation of a web browser \cite{javaselejavaselenium}. It can emulate user interaction with browsers to perform any web browsing task as a human usually do. The core of Selenium is WebDriver \cite{webdriver} which is an interface to write instruction sets that can run on many browsers such as Chrome, Firefox, or Edge. Selenium support multiple programming languages such as Java, Python, C Sharp
The component design of Selenium can be illustrated using the below diagram. 

WebDrivers communicate with the browser through a driver. WebDriver passes and receives information via the same route. Using Selenium, we can automate the browser to click on each link within the web page and search for the keywords. The searching and content matching would be repeated until the Selenium reached the end of the website.

\begin{figure}[h]
    \centering
    \includegraphics[width=0.4\paperwidth]{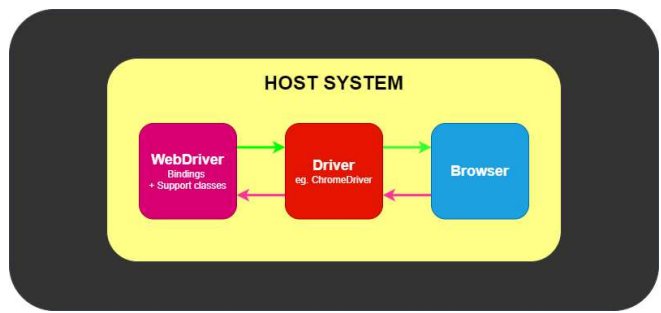}
    \caption{Selenium webdriver }
    \label{fig:my_label}
\end{figure}

\paragraph{Namecheap Cloud server }
Namecheap cloud server is a cloud service that we would use to deploy our backend and front-end applications and have it constantly running with no downtime. Namecheap hosting provides a versatile configuration option that supports multi-programming language applications, web API, authentication schemes, and schemeless databases, suitable for my project. 

\paragraph{VueJS}
VueJs is a javascript front end framework \cite{vuejs} that supports material design. We can incorporate the front-end UI with a Java backend server using this framework. Data requests can be made to the backend server using the REST API library such as Axios. In addition, VueJS can be extended by using Vuetify \cite{vuetify} to develop a clean, simple but effective UI. Vuetify framework provided a set of pre-designed layouts and web components to develop a prototype quickly. The reactive mechanism would allow us to render the data in real-time to give users an instantaneous result if the scanned website is simple enough.

\subsection{Preliminary Design And Actual Implementation Comparison  }
\paragraph{CCPA Informing Page}
In the design, we planned to implement a feature that can present actual content of CCPA to consumers when they first land on our website. This feature could be concluded by constructing a simple text web page to show the CCPA act. To aid the presentation, we would use a simple structure with formatted text to highlight critical information from the act. 
In the implementation, we have used the modern UI framework VueJS and a single web page template to make the content more engaging. In addition, we utilize UI blocks and cards to highlight important information and support assistive technology. We also embed a link that redirects to the CCPA official page for comprehensibility. Overall, we rate the implementation of this feature as completed with high satisfaction. 

\paragraph{CCPA Compliance Verifier}
In application design, this is the main feature of the application. The value of this feature is to provide users with a swift and effortless way to confirm CCPA on a website. Principally, the tool will check if a website mentions CCPA and provide sufficient rights to consumers over their data. In addition, the tool's user interface will have a field for users to enter the company website URL. 
In the implementation, we extend this feature by providing consumers with two ways to check a website, choose a website from a list, or enter the URL manually. In addition, consumers can access the CCPA verification easily by clicking on UI card elements. However, because this application feature sometimes suffers from false results, we only rate the implementation of this feature as complete with reasonable satisfaction. Therefore, we will keep improving the performance of this feature in the future.

\paragraph{CCPA Compliance Helper}
In the application design, the CCPA Compliance Helper is an extended feature out of the scope to assist companies with CCPA complying. Based on the CCPA verifying result, this feature proposes adjustments to make the website fully comply with CCPA.

However, this feature has not been completed in the current application version because of time constraints and dependency issues. Therefore, we will finish this feature by the next developmental stage.

\section{\centering CCPA Verifier - Feature Flow}
\label{sec:testing}
The following is the flow of the CCPA verifier feature. This section will present what a consumer has to do to verify CCPA on a website using our application. Again, the application is fully functional at www.ccpaverifier.com.  

First, after entering the homepage, consumers click on the feature button to go to the application feature page. 
\begin{figure}[H]
    \centering
    \includegraphics[width=0.4\paperwidth]{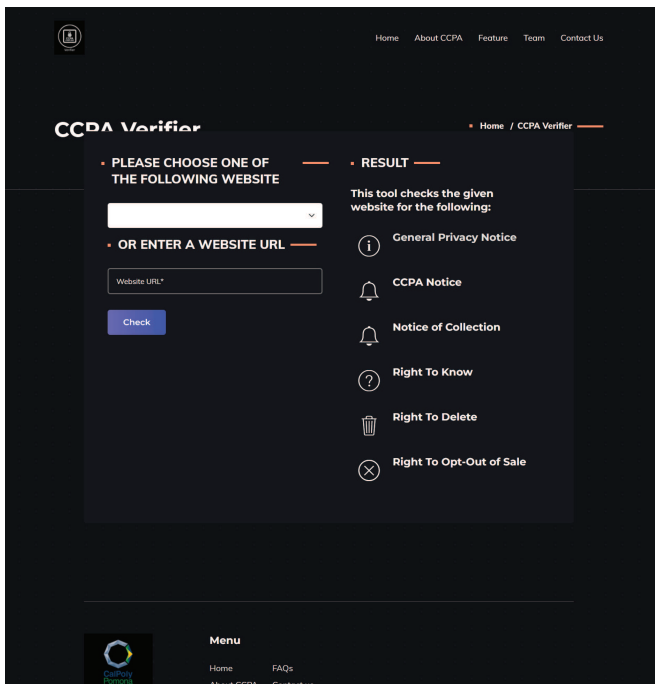}
    \caption{CCPA verifier page}
    \label{fig:my_label}
\end{figure}
Here, consumers will have two options to use this feature. First, consumers can choose a website from the drop-down list or enter the URL manually in the typing box. Then, after selecting, consumers press enter or click the check button to execute a CCPA verifier feature. 
\begin{figure}[H]
    \centering
    \includegraphics[width=0.4\paperwidth]{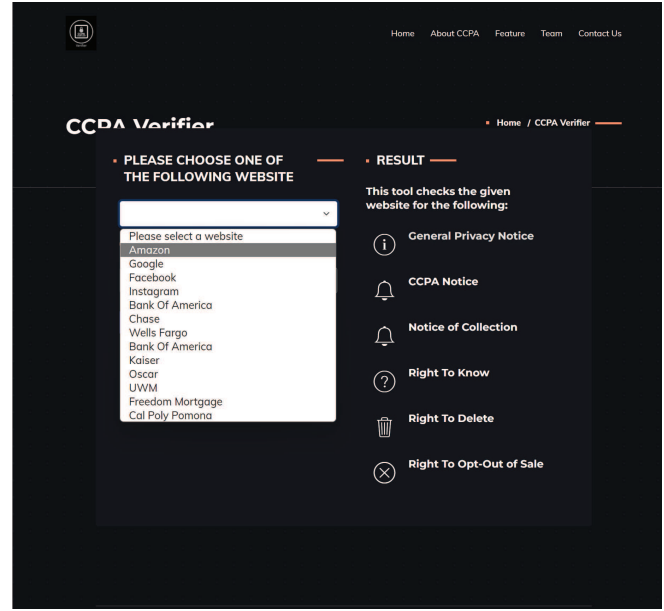}
    \caption{Website selection for CCPA verifying}
    \label{fig:my_label}
\end{figure}
As soon as consumers click the check button, the application will start CCPA verifying process. It takes approximately thirty seconds to one minute to receive a result. The pending state is as shown in the below figure. In this state, the front-end application will send a REST request which includes a website URL. Then the back-end will call initialize a web crawler using selenium driver to crawl data from the target website. As soon as a result is generated, it will be sent back to the UI to be rendered. 

\begin{figure}[H]
    \centering
    \includegraphics[width=0.4\paperwidth]{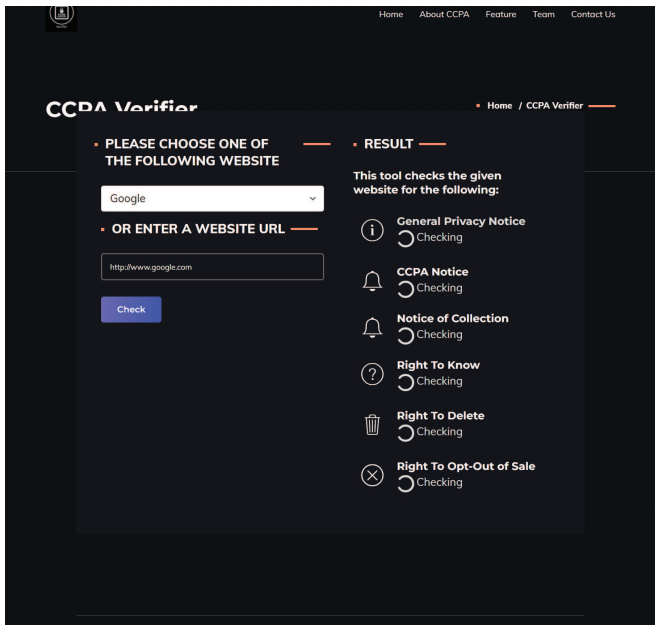}
    \caption{CCPA verifying processing is being performed}
    \label{fig:my_label}
\end{figure}
When a CCPA verifier process is finished with a success or error, if successful, the back end server will return a message that contains a collection of a result of  Boolean values. The boolean values are indicators that check if a website has a general privacy notice, CCPA notice, notice of collection, right to know, right to Delete, and right to opt-out of sale. The front-end UI will render the result based on the Boolean values in the response. The example is as follows: 

\begin{figure}[H]
    \centering
    \includegraphics[width=0.4\paperwidth]{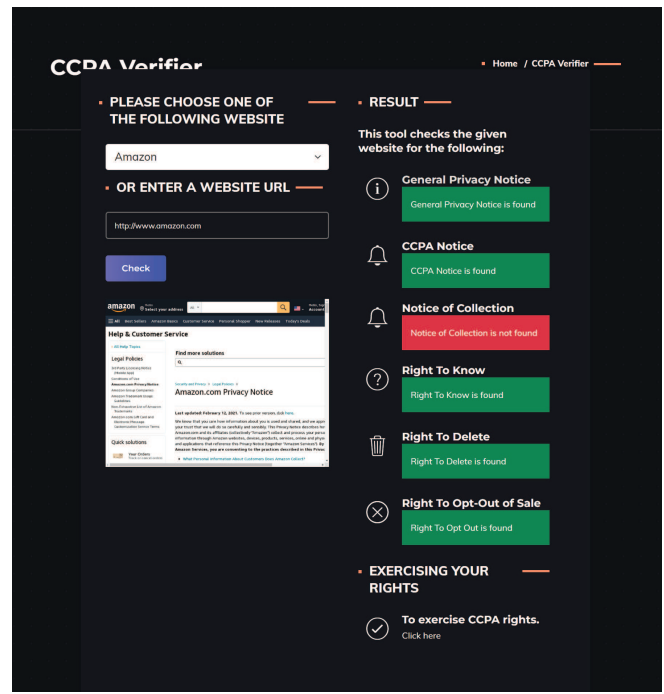}
    \caption{CCPA verifying is successful}
    \label{fig:my_label}
\end{figure}
If the verifier detects CCPA compliance on a website, it will provide a link for consumers to exercise CCPA rights. That link will redirect consumers to the CCPA page of the tested website. Consumers can submit requests to access, delete data, or opt out of sale. This feature will save consumers significant time by letting consumers access the web page where they submit CCPA requests quickly. In another case,  if the application returns a result that indicates the tested website does not comply with CCPA, the application will provide instructions for consumers to file a complaint to the Office of Attorney General. In addition to the instruction, consumers can download a screenshot of the result to file a complaint.

\begin{figure}[H]
    \centering
    \includegraphics[width=0.4\paperwidth]{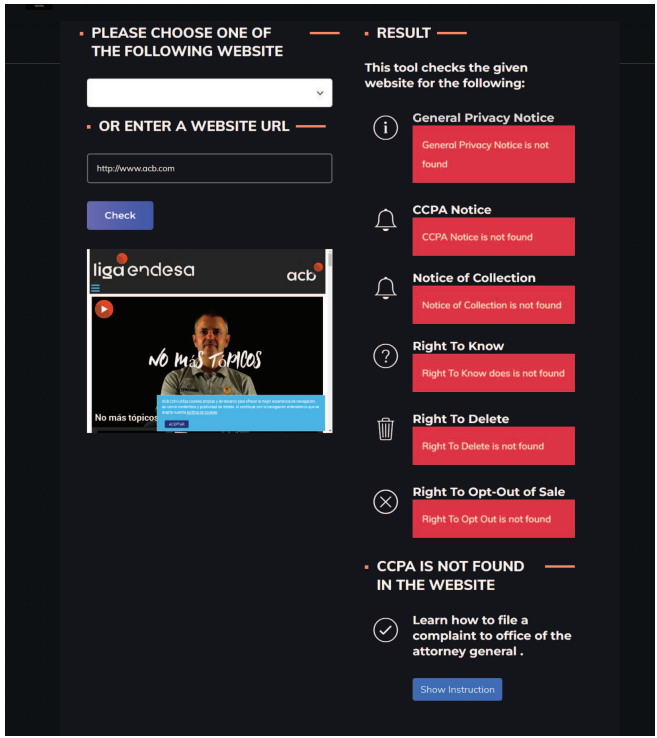}
    \caption{CCPA is not found on the website }
    \label{fig:my_label}
\end{figure}

\begin{figure}[H]
    \centering
    \includegraphics[width=0.4\paperwidth]{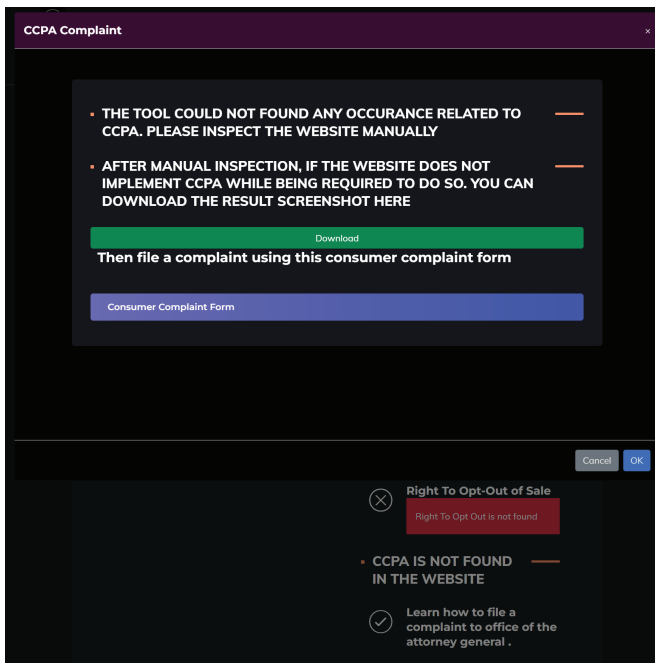}
    \caption{Instruction of filing a complaint }
    \label{fig:my_label}
\end{figure}

If the verifier fails when checking, it will present an error to inform the consumers what happened with the process. 

\begin{figure}[H]
    \centering
    \includegraphics[width=0.4\paperwidth]{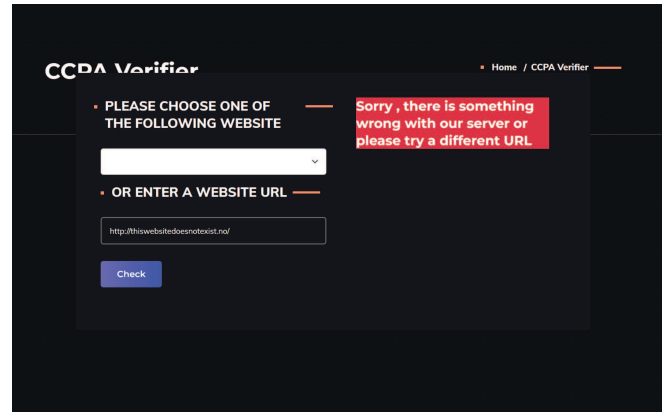}
    \caption{CCPA verifying is successful}
    \label{fig:my_label}
\end{figure}

\section{\centering Evaluation of Results}
\label{sec:evaluation}

We can evaluate the effectiveness of the CCPA checking tool by comparing the tool's result with a table of ground truth values. Ground truth values are the CCPA compliance result obtained by manually reviewing CCPA notices on websites. First, we picked popular websites to check their CCPA compliance manually and then constructed a criteria table in section 2.1. In the next step, we would use the CCPA tool to inspect the websites to retrieve the results about their CCPA compliance automatically. After that, we compared the application's result with the ground truth table to measure the tool's accuracy. The websites included in the test were the same websites in the literature review section. Finally, we tested the tool with websites implemented by different web technologies for generalization.

The evaluation results showed that the CCPA verifier worked well with most of the websites on the list. However, the application got some difficulties navigating the CCPA notice page on some websites, such as Chase's website. Still, we successfully implemented extended functions to guide the crawler to reach the CCPA site partially. After that, the application was able to return a result as usual. Overall, the application retrieved almost one hundred percent accuracy when dealing with the websites from the list, thanks to the extended functions that make the crawler work specifically with a website. For the websites that are outside of the list, we still experienced hit and miss. Sometimes, the application couldn't detect some CCPA implementation on some websites, but overall, the application performed well in detecting CCPA privacy pages. To increase the performance of the websites in the lists, we keep adding more boundary checks to improve the generalization of the application to make it work with more websites.

Additionally, we evaluated the robustness of the application based on the functionalities and the efficiency of the user interface. The application user interface should be straightforward to provide results with minimal effort. For example, the application should not ask a user to create an account to use. Furthermore, we enforce an uncluttered and intuitive UI design that aid the result presentation. Finally, we let Cal Poly Pomona students use the tool and then provide feedback to evaluate the effortlessness of usage. Generally, the student's feedbacks are positive. Most of them said the application's UI is easy to understand, and they could access the feature easily. The application's responding time is reasonable. The UI outline and components only need minor improvement to present the contents better. Some button hyperlinks are required to be fixed. Overall, the application's UI satisfies the business requirement mentioned earlier.

\subsection{False Positive and False Negative Issues}
Fundamentally, the business logic behind the CCPA verifier function utilizes web crawling and natural language processing. To determine if a website complies with CCPA, our application needs to detect a required number of keywords that match a set of  CCPA criteria. If a website does not implement  CCPA but mentions the right keywords, the verifier will make a false assumption. Therefore, our application logic would falsely determine that website complies with one or more CCPA criteria. On the other hand, our application will also face some false-negative cases.\\
Another issue comes from the sensitive keyword matching algorithm that our application is powered by. We have a bucket of keywords required to match each CCPA criteria. However, suppose a website uses keywords for  CCPA implementation which are different from our keywords. Again, our application would make false assumptions and then conclude the website does not comply with CCPA. In short, the false positive and negative issue depends on how we define boundaries for the keywords. 
The figure below is an example of a false positive case our application faced. We used our CCPA verifier to test our website, ccpaverifier.com, and it returned a wrong result. We do not have CCPA implementation on our website in this scenario, but the application logic still detected that our website complies with some CCPA criteria. However, because our website has many keywords about CCPA, it makes the CCPA verifier misunderstand that our website was implementing CCPA.

\begin{figure}[H]
    \centering
    \includegraphics[width=0.4\textwidth]{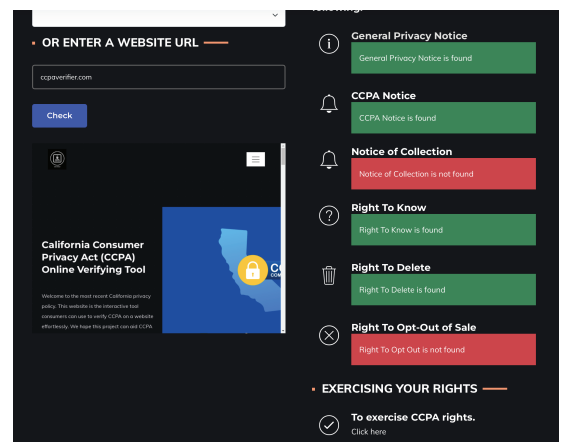}
    \caption{An example of CCPA false positive}
    \label{fig:my_label}
\end{figure}

Consequently, we believe in limiting these false positive and negative issues; we will keep improving our back-end logic to make the web crawler better navigate through websites. If the crawler can navigate correctly to the privacy pages rather than elsewhere, it will likely come up with the correct determination. In addition, understanding a website context is more valuable than checking exact keywords. If we can find an approach to make, our application recognizes that if a website mentions CCPA by extracting the content, we will get much more accurate results. To do that, an AI technique, natural language processing (NLP), is what we should look at for our future research. NLP will allow our application to analyze and understand the context of a document and then decide if that document is relevant to a topic based on a confidence score. Therefore, the general concept seems to be ideal for our application. It can be in our next developmental stage in the future. 

\subsection{Accessibility Support}
We also enhanced our website to make it compatible with assistive technology for disabled people. We have to make necessary changes to the colors and layout of our application UI. To know which web application components we should revise, we evaluated accessibility support using Screen Reader - Assistive Technology: NVDA \cite{NVDA}, WAVE Web Accessibility Evaluator Tool \cite{WAVE}, Color Contrast Analyzer \cite{Color-Contrast-Analyzer}. The result that these programs generated provided us with suggestions for improving our website interface. 

\begin{itemize}
    \item \textbf{Visual focus}: Used by sighted keyboard-only users to identify their location on the page. The tool found on our website has a low visual focus which means users will have difficulty seeing which element has focus. For example, when we hover the mouse on the feature button, read more button, back to top button, the focus is not visible enough for people who have vision-impairing to see. 
    \item \textbf{Keyboard Operable}: When collapsing the browser's window into a smaller window, the navigation menu collapses into a hamburger button. Consumers cannot use either keyboard and mouse to navigate the web page.
    \item \textbf{Announcement}
        \begin{itemize}
            \item Linkedin buttons are announced as lists when they are buttons, and the link destination is not announced
            \item Text lists are not formatted as lists. Assistive technology users will have difficulty identifying the text is related

            \begin{figure}[H]
                \centering
                \includegraphics[width=0.4\textwidth]{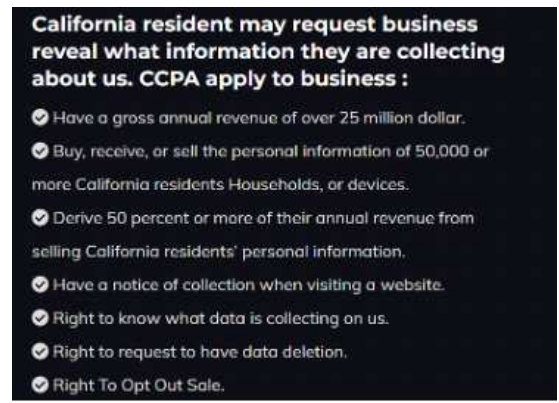}
                \caption{Text list is not compatible with assistive technology}
                \label{fig:my_label}
            \end{figure}

        \end{itemize}
        
    \item \textbf{Link}: Links open in new window without advanced warning. Assistive technology users will have difficulty navigating back to the website (can be fixed with title attribute for the link in the code)
    \begin{figure}[H]
        \centering
        \includegraphics{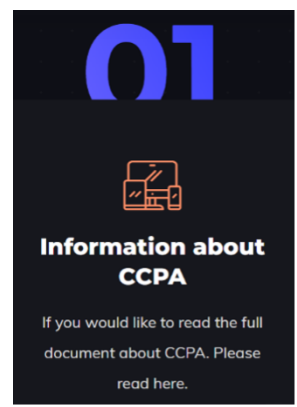}
        \caption{A link issue }
        \label{fig:my_label}
    \end{figure}
    \item \textbf{Headings }: Assistive technology users often navigate the page with headings
        \begin{itemize}
            \item Headings don’t have a logical order.
                \begin{figure}[H]
                    \centering
                    \includegraphics[width=0.4\paperwidth]{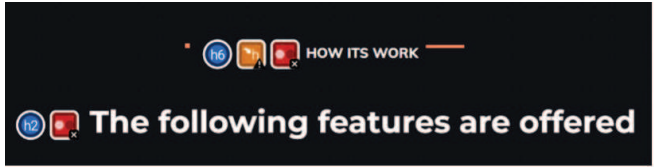}
                    \caption{Headings do not have a logical order}
                    \label{fig:my_label}
                \end{figure}
            \item Heading levels are skipped
                \begin{figure}[H]
                    \centering
                    \includegraphics{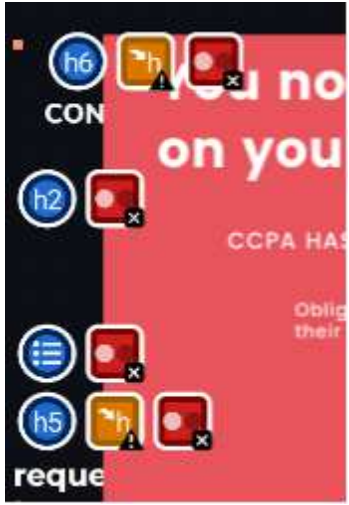}
                    \caption{Heading levels are skipped}
                    \label{fig:my_label}
                \end{figure}
        \end{itemize}
    \item \textbf{Alt text }:  used to describe images to assistive technology users
        \begin{itemize}
            \item First image on home page contains text. Assistive technology users cannot determine the purpose of the image (if the image is decorative, can set alt to decorative (alt=””)) Note: If the image relays the same information as a text paragraph on the website, We should mark it as decorative since it’s a visual representation of the information. 
          
            \begin{figure}[H]
                \centering
                \includegraphics[width=0.4\paperwidth]{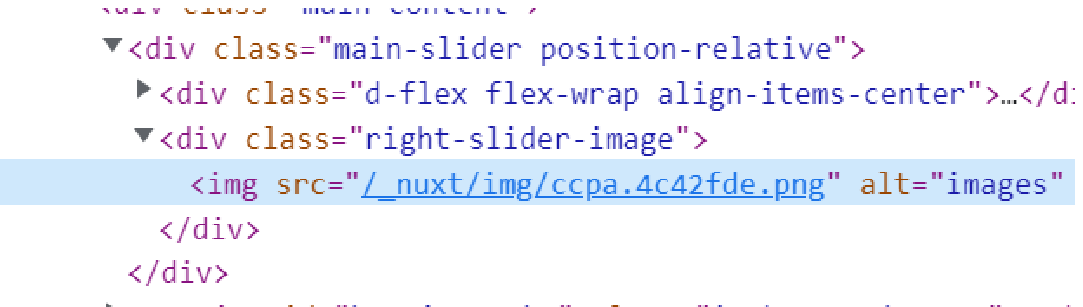}
                \caption{Alt text issue}
                \label{fig:my_label}
            \end{figure}
       
        \end{itemize}
\end{itemize}

Based on the accessibility report, we began to adjust our website layout and design to be better compatible with assistive technology. Upon completing the adjustment, people with disabilities can use assistive technology to use our website. We believe it is crucial to let disabled people access our application to be aware of the rights that CCPA provides. Regardless of people's conditions, consumers get affected in the same way in term of data collection and sharing made by organizations 

\section{Future Works}
For future works, we may apply the technique that the Cookiebot application \cite{cookiebot} is using to scan the cookie to understand which parties a website shares data with. Hence, in the next version of the application, we will deploy a feature that allows our application to detect the cookies being used on a website. Especially some websites now offer users an option to opt-out of being tracked, which means those websites should not install tracking cookies in the consumer's browser if consumers choose to opt out. Therefore, the scanning cookies future will evaluate the website's cookie status before and after the opt-out of being tracking request. That functionality will validate the trustworthy of a website toward the opt-out of being tracked offer. 

The second feature we would like to develop next is data leaking incident searching. In this feature, when performing CCPA verification, our application will also connect to a search engine such as Google search to look for data leaks, data breaches, or data scandals related to the company being tested. Then that information will be presented to consumers to make consumers more aware of the company's data protection practices. Consumers will also use this future multiple times to monitor the company's activities to see if they improve their privacy policy after the incident occurs. The implementation of this is simple; we will display an additional card next to the CCPA criteria result in that displays the top 5 of the results after searching for the related data incidents.

We believe these two future features would be essential addons to our application by making it more interactive and informational and bringing more consumer awareness about data privacy. 

\section{Conclusion}
In conclusion, many California organizations have implemented CCPA and established platforms for consumers to manage their data. CCPA can be applied to most organizations that collect and use consumer data, except financial institutions and healthcare providers. These types of businesses have other privacy acts to exempt them from CCPA. However, these privacy acts cannot protect organizations from being sued if data breaches happen, they have to take full responsibility for being neglected in securing consumers' data. Honoring CCPA would allow companies to give consumers more control over their data. Consumers can decide the best approach to manage their data by accessing and deleting them. \\
We completed this project by finishing an evaluation of the organization's CCPA implementation and a web application that aids with the CCPA. We believe both the assessment and the application would benefit California consumers by helping them to exercise CCPA data rights. In addition, organizations also benefit from this project by having an efficient tool to do self-audit to adjust privacy policy and then comply with the newest data privacy law from California.\\
Finally, we will keep maintaining our application's server opening for the public to access. We also monitor the application's performance and improve it to produce a result more quickly and accurately in the next versions.

\bibliographystyle{IEEEtran}
\bibliography{IEEEabrv,biblio.bib}

\end{document}